\documentclass[a4paper]{JHEP3}
\usepackage{epsfig,bbm}

\usepackage{ifthen}
\usepackage{epsfig}
\usepackage{amssymb}
\usepackage{graphicx}
\usepackage{amsmath}
\usepackage[center]{subfigure}
\renewcommand{\text}[1]{%
\ifthenelse{\equal{#1}{mB}}{m_B}{}%
\ifthenelse{\equal{#1}{sig}}{\sigma}{}%
\ifthenelse{\equal{#1}{mc}}{m_c}{}%
\ifthenelse{\equal{#1}{q2}}{q^2}{}%
\ifthenelse{\equal{#1}{om}}{\omega}{}%
}
\newcommand{\ba}{\begin{eqnarray}}
\newcommand{\ea}{\end{eqnarray}}
\newcommand{\be}{\begin{equation}}
\newcommand{\ee}{\end{equation}}

\newcommand{\A}{{\cal A}}

\newcommand{\DS}[1]{/\!\!\!#1}

\title{\Large \bf $B \to K \ell^{+}\ell^{-}$ decay at large hadronic recoil}

\author{A.~Khodjamirian, Th.~Mannel and Y.-M.~Wang\footnote
{Address after September 30, 2012: Physik Department T31, James-Franck-Stra${\ss}$e,
Technische Universit\"{a}t M\"{u}nchen, D-85748 Garching, Germany.} \\
\it  Theoretische Physik 1, Naturwissenschaftlich-Technische Fakult\"at, Universit\"at
Siegen, \\ D-57068 Siegen, Germany }

\abstract{
We predict the amplitude
of  the $B\to K \ell^+\ell^-$  decay in
the region of the dilepton invariant mass  squared $0<q^2\leq m_{J/\psi}^2$,
that is, at large hadronic recoil.
The $B\to K$ form factors  entering the factorizable
part of the decay amplitude are obtained from QCD light-cone
sum rules.  The nonlocal effects, generated by
the four-quark and penguin operators combined with the
electromagnetic interaction, are calculated at $q^2<0$,
far below the hadronic thresholds. For hard-gluon contributions
we employ the QCD factorization approach.
The soft-gluon nonfactorizable contributions
are estimated from QCD light-cone sum rules.
The result  of  the  calculation is matched to the hadronic
dispersion relation in the variable $q^2$, which is then continued  to the
kinematical region of the decay.
The overall effect of nonlocal contributions in $B\to K\ell^+\ell^-$
at large hadronic recoil is moderate.
The main uncertainty of the predicted $B\to K  \ell^+\ell^-$ partial width is caused
by the $B\to K$ form factors.  Furthermore, the isospin asymmetry in this decay
is expected to be very small. We  investigate the
deviation of the observables from the Standard Model predictions
by introducing a generic new physics contribution to the effective
Hamiltonian. }


\begin{document}

\section{Introduction}

After a very successful start, the LHCb experiment has already
provided new measurements of  exclusive flavour-changing neutral
current (FCNC) decays of $B$ mesons \cite{LHCbFBA}-\cite{LHCbnew},
continuing the studies of these  decays
carried out at $B$-factories \cite{Wei:2009zv,Lees:2012vw}
and Tevatron \cite{Aaltonen:2011qs}.
The favourite decay channel is $B\to K^*\ell^+\ell^-$
where a rich kinematics allows one to measure several
nontrivial observables sensitive to the underlying  FCNC $b\to s \ell^+\ell^-$
transitions. So far, the measurements of
these decays did not reveal any significant
deviation from Standard Model (SM). In particular,  the zero crossing point  of  the forward-backward
asymmetry in $B\to K^*\ell^+\ell^-$ predicted in SM has recently been observed
\cite{LHCbFBA} within the expected interval of the variable $q^2$,
the invariant mass squared of the lepton pair.
Meanwhile, the measured isospin asymmetry in $B\to K^{(*)}\ell^+\ell^-$,
allows for the  values larger than expected in SM,
hence, this observable deserves a further careful study.

The theory of rare semileptonic decays,
such as $B\to K^{(*)}\ell^+\ell^-$, suffers from hadronic uncertainties.
Apart from the heavy-to-light form factors, determining the
contributions of the leading FCNC operators in the effective
Hamiltonian, the decay amplitudes  receive contributions
of the current-current and penguin operators  combined with the  lepton-pair emission
via electromagnetic (e.m.) interaction.
A common feature of all these effects is that they are {\em nonlocal} because
the quark-flavour transition is separated from the virtual-photon emission,
and the characteristic distances of this separation are not necessarily small.
The most important nonlocal contribution is generated by the current-current
operators with $c$ quarks,  forming ``charm-loops'' after photon emission.
The charm loops  turn into intermediate charmonium states at $q^2\geq m_{J/\psi}^2$.
The nonlocal effects in $B\to K^{(*)}\ell^+\ell^-$ have been estimated in the literature, within
different models and approximations. The most complete
analysis at large hadronic recoil (low $q^2$)
was done in the QCD-factorization framework in \cite{BFS}.
In that approach, the quark-gluon
diagrams of various factorizable and nonfactorizable nonlocal contributions
were calculated at timelike $q^2$, below  charmonium
thresholds.

In our previous paper \cite{KMPW} the soft-gluon emission from the charm loop
in $B\to K^{(*)}\ell^+\ell^-$ -- an effect not accessible in QCD factorization approach--
was estimated.  To that end, the light-cone
operator product expansion (OPE) at $q^2\ll 4m_c^2$ was employed and the
soft-gluon emission was effectively resummed in a
nonlocal quark-antiquark-gluon operator. The resulting hadronic matrix elements
were then estimated using  QCD light-cone sum rules (LCSR),  i.e., the same approach
which is used to calculate the $B\to K^{(*)}$ form factors.
Being suppressed by the powers of $\Lambda_{QCD}^2/(4m_c^2-q^2)$,
the soft-gluon contribution
to the decay amplitude grows at $q^2$ approaching the $\bar{c}c$ threshold,
To avoid the divergence at
$q^2 \simeq 4m_c^2 $ ,
the hadronic dispersion relation in the $q^2$ channel was employed
in  \cite{KMPW} and matched to the calculation  result at $q^2\ll 4m_c^2$.

The aim of this paper is to perform a complete
account of hadronic nonlocal  effects for one particular channel $B\to K\ell^+\ell^-$.
We will apply the same approach as
in \cite{KMPW}, this time including   the complete effective
Hamiltonian.
In this case,  light-quark loops also contribute to the amplitude,
therefore we  calculate the nonlocal effects in QCD  in the spacelike region $q^2<0$,
sufficiently far from all quark-antiquark thresholds.
Soft-gluon emission evaluated in \cite{KMPW} is
taken into account for  quark loops with different flavours. In addition,
we calculate a new effect
of soft-gluon emission from the gluon-penguin operator
applying the LCSR approach. We also take into account the hard-gluon NLO contributions.
In  the correlation functions used to derive LCSR's,
the hard-gluon exchanges generate multiloop and multiscale diagrams
which demand dedicated calculational efforts that are far beyond our scope.
Instead, we approximate the hard-gluon nonlocal effects, employing
the QCD factorization approach \cite{BFS} at $q^2<0$.
Following \cite{KMPW}, the physical region in $B\to K \ell^+\ell^-$
is accessed  via hadronic dispersion relations in the variable $q^2$.
In these relations, in addition to the charmonium states, the light vector
mesons contribute. Finally, we predict  the observables for $B\to K \ell^+\ell^-$
including the differential width  and isospin asymmetry. Our results are applicable
at $q^2\leq m_{J/\psi}^2$ where they are compared with the available experimental data.

The $B\to K\ell^+\ell^-$ decay channel chosen here
has less observables and a smaller branching fraction than $B\to K^*\ell^+\ell^-$.
In turn, the relative simplicity makes the kaon mode a more convenient
study object for hadronic effects. First of all, the current  accuracy
of the $B\to K$ form factors is better than
for $B\to K^*$ form factors. Note that the latter form factors are available only in
the quenched approximation of the lattice QCD. The $B\to K$ form factors
used here are calculated from LCSR with kaon distribution amplitudes (DA's).
These sum rules are  free from the ``systematic'' uncertainty
of LCSR's for $B\to K^*$ form factors caused by
neglecting the $K^*\to K\pi$ width.  The alternative LCSR's
with $B$-meson DA's  implicitly overcome this problem if one relies on the
duality approximation in the $K^*$ channel.
However, the  $B$-meson DA's  still suffers from large  uncertainties of their parameters,
such as the inverse moment. Since the form factors not only determine the leading
operator contributions but also enter the factorizable
nonlocal effects, we expect that the hadronic input
in $B\to K \ell^+\ell^-$ is currently under a better control than in $B\to K^* \ell^+\ell^-$.
This circumstance strengthens the case for using
the kaon channel  as a perspective
tool to trace generic new physics (see e.g., \cite{Bobeth:2007dw,Buchalla_etal}).

The plan of the paper is as follows.
In Section~2 we define the nonlocal hadronic matrix elements entering the $B\to K \ell^+\ell^-$
amplitude. The nonlocal effects are listed in terms of quark-gluon diagrams.
In Section~3 we present the corresponding hadronic matrix elements in the spacelike region obtained by combining LCSR's with the  QCD factorization
approach. In Section~4 the LCSR for the soft contribution of the gluon-penguin operator is derived. Furthermore,
in Section~5 we  specify the use of hadronic dispersion relations to access
the physical region.
In Section 6 the relevant numerical analysis is performed.  The resulting
predictions for the $B\to K\ell^+\ell^-$ differential
width and isospin asymmetry are presented in Section~7 where we
also discuss the influence of generic new physics on these observables.
Section~8 is reserved for the concluding discussion and outlook.
In Appendix~A, the operators
of the effective Hamiltonian and their Wilson coefficients are specified, Appendix~B
contains the definitions of $B$-meson DA's and  Appendix~C
the expressions related to the LCSR derivation.

\section{ Hadronic effects in the decay amplitude}

The $B \to K \ell^+\ell^-$ decay amplitude
in SM  is  given by the  matrix element  of the $b\to s \ell^+\ell^-$   effective Hamiltonian \cite{Heff,BBL}:
\be
H_{eff}= -\frac{4G_{F}}{\sqrt{2}}V_{tb}V_{ts}^{*}
{\sum\limits_{i=1}^{10}} C_{i}({\mu}) O_{i}({\mu})\,.
\label{eq:Heff}
\ee
The effective local operators $O_i$  and the numerical values of their Wilson coefficients
$C_i$ are presented in App.~A. In this paper we generally neglect the CKM-suppressed contributions proportional to $V_{ub}V_{us}^*$, retaining them only in the calculation of
the isospin asymmetry where their numerical impact is  non-negligible.

\begin{figure}[tb]
\begin{center}
\hspace{-1 cm}
\includegraphics[scale=0.6]{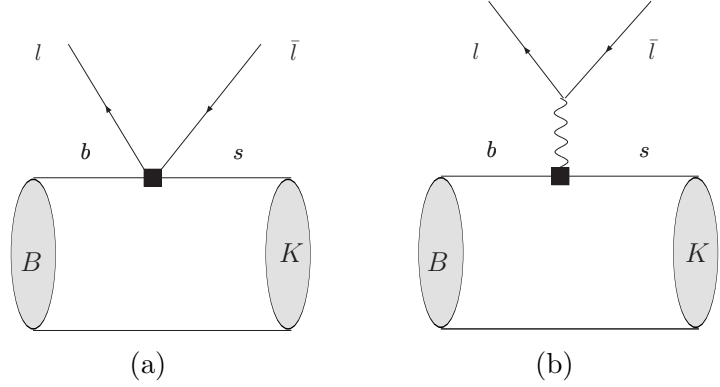} \\
{} \hspace{-1.2cm}(a)\hspace{4.9cm}(b)
\caption{\it  FCNC contributions to $B\to K \ell^+\ell^-$
due to the  effective operators $O_{\,9,10}$ (a)
and $O_7$ (b) denoted as black squares.}
\label{fig:dom}
\end{center}
\end{figure}

In the decay  amplitude  we separate the dominant
factorizable contributions of the operators $O_{7,9,10}$   (Fig.~\ref{fig:dom}):
\begin{eqnarray}
A (B \to K \ell^{+} \ell^{-}) =
-\langle K (p)\ell^+\ell^-\mid H_{eff}\mid B(p+q)\rangle \nonumber\\
={ G_F \over  \sqrt{2} } {\alpha_{em}
\over \pi} V_{tb} V_{ts}^{\ast} \Bigg[
\bar{\ell}\gamma_{\mu} \ell\, p^\mu\bigg( C_9 f^{+}_{BK}(q^2)
+ {2 (m_b+m_s) \over m_B+m_K} C_7^{eff} f^{T}_{BK}(q^2) \bigg)
\nonumber\\
+
\bar{\ell} \gamma_{\mu} \gamma_5 \ell \, p^\mu C_{10}  f^{+}_{BK}(q^2)
-(16\pi^2)\frac{\bar{\ell}\gamma_\mu\ell}{q^2}{\cal H}^{(BK)}_\mu
\bigg]\,,
\label{eq:ampl}
\end{eqnarray}
where  $f^{+(T)}_{BK}(q^2)$ is the usual $B\to K$
vector (tensor) form factor and, according to \cite{BBL},
$C_7^{eff}=C_7-{1 \over 3} C_5-C_6$.
The nonlocal hadronic matrix element
\ba {\cal H}^{(BK)}_\mu= i\int d^4xe^{iq\cdot x} \langle
K(p)| T \Big\{j^{em}_\mu(x), \! \Big[ C_1O_1^{(c)}(0)+
C_2O_2^{(c)}(0)\nonumber \\
+\sum\limits _{k=3-6,8g} C_kO_k(0) \Big]\Big\}|
B(p+q) \rangle
= [(p\cdot q) q^\mu - q^2 p^\mu]{\cal H} ^{(BK)}(q^2)
\label{eq:matr}
\ea
contains the contributions of
all remaining operators in (\ref{eq:Heff}) combined  with the quark electromagnetic (e.m.)
current $j^{em}_\mu=\sum_{q=u,d,s,c,b}Q_q\,\bar{q}\gamma_\mu q $.  Our definition
of ${\cal H} ^{(BK)}$ differs from the one used in \cite{KMPW} by including
the quark-charge factors $Q_q$. Furthermore, due to renormalization
effects, one has to  replace \cite{BBL} above
$C_{8g} \!\! \to \! C_8^{eff}= C_{8g} + (4 C_3-C_5)/3$.
After substituting  (\ref{eq:matr})  in (\ref{eq:ampl}), due to
conservation of the leptonic current, only the structure $q^2 p^\mu$
remains in (\ref{eq:matr}), canceling  the photon propagator
$1/q^2$. Hence, opposite to the case of $B\to K^*\ell^+\ell^-$,
there is no  kinematic enhancement of the $B\to K \ell^+\ell^-$ amplitude
at low $q^2$.  Dividing the
invariant amplitude ${\cal H}^{(BK)}(q^2) $
by the form factor $f^{+}_{BK}(q^2)$,  it is convenient to represent the nonlocal
hadronic effect  in a form of  a (process- and $q^2$-dependent) correction
to the  coefficient $C_9$,
\be
\Delta C_9^{(BK)}(q^2)=\frac{16\pi^2{\cal H} ^{(BK)}(q^2) }{f^+_{BK}(q^2)}\,.
\label{eq:deltc9}
\ee

\begin{figure}[tb]
\begin{center}
\hspace{-1 cm}
\includegraphics[scale=0.6]{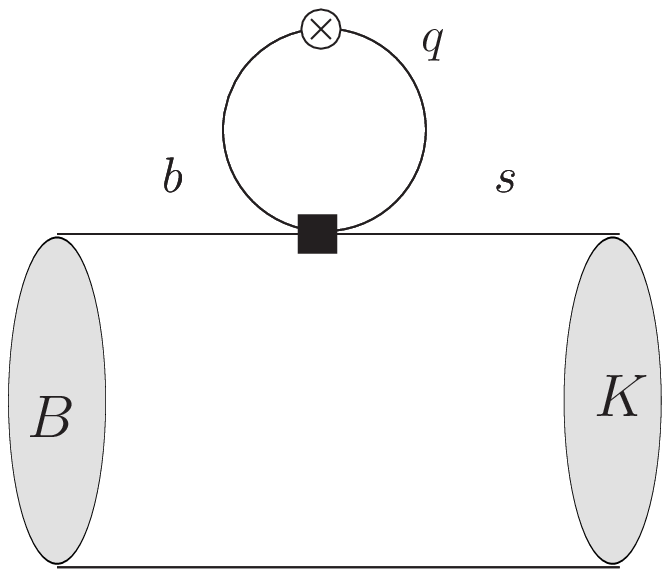} ~~
\includegraphics[scale=0.6]{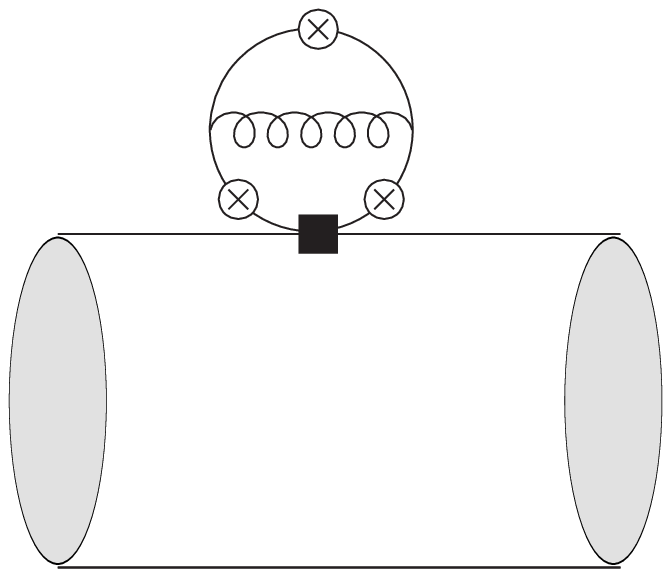}~~
\includegraphics[scale=0.6]{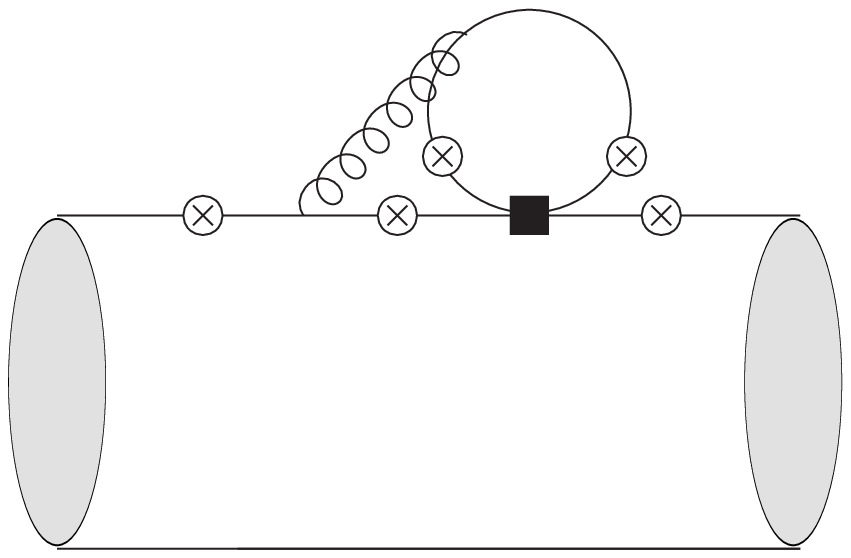}\\
{} \hspace{-1.2cm} (a) \hspace{3.5cm}(b)\hspace{4.8cm}(c)
\caption{\it Factorizable quark-loop contributions to $B\to K \ell^+\ell^-$
due to four-quark effective operators $O_{1,2}^{c}$ and $O_{3-6}$.
Crossed circles denote the possible points of the virtual photon emission. Diagrams
similar to (c) with the gluon line attached to the $s$-quark line are not shown.}
\label{fig:fact}
\end{center}
\end{figure}

\begin{figure}[tb]
\begin{center}
\hspace{-1 cm}
\includegraphics[scale=0.6]{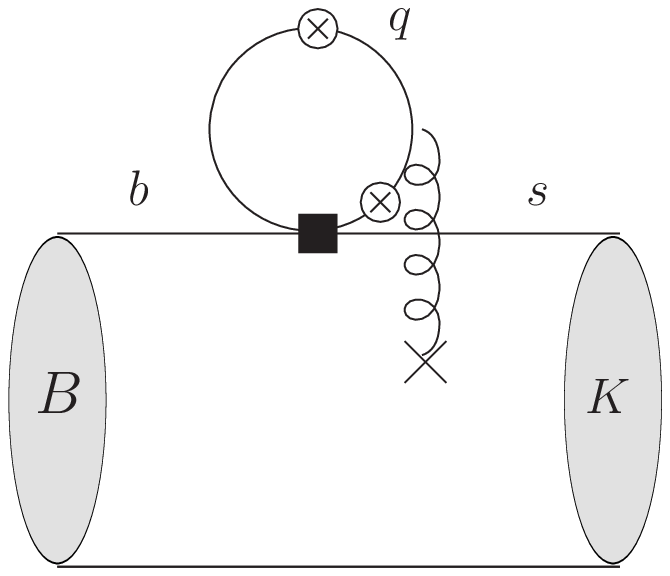}~~
\includegraphics[scale=0.6]{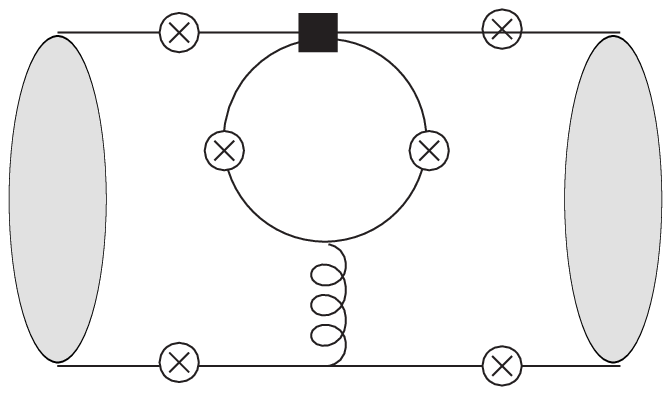}\\
{}\hspace{-0.5cm}(a) \hspace{4.0cm}(b)
\caption{ \it Nonfactorizable quark-loop contributions to $B\to K \ell^+\ell^-$: (a) with soft gluon (denoted by
crossed line) and (b) with hard-gluon.}
\label{fig:nonf}
\end{center}
\end{figure}

\begin{figure}[t]
\begin{center}
\hspace{-1 cm}
\includegraphics[scale=0.6]{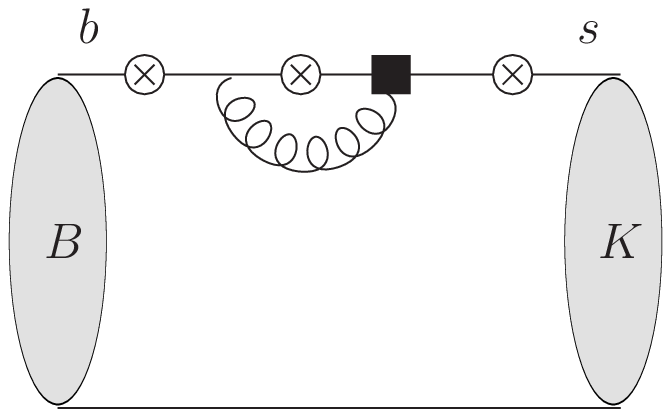}~~
\includegraphics[scale=0.6]{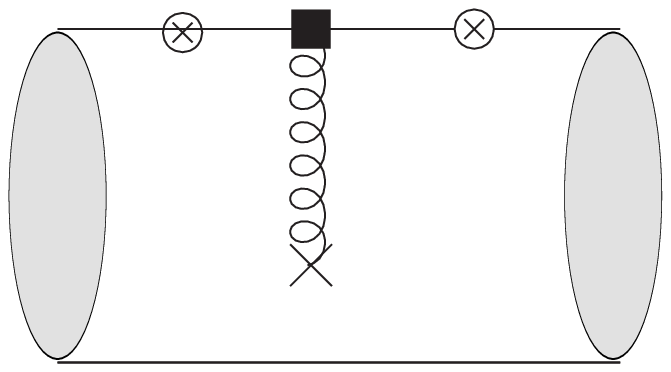}~~
\includegraphics[scale=0.6]{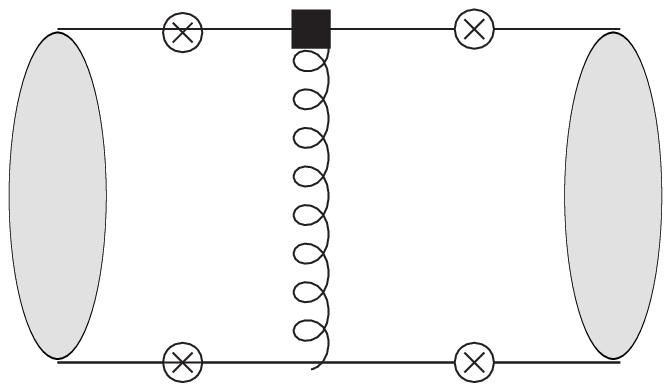}\\
 (a) \hspace{4.5cm}(b)\hspace{4.8cm}(c)
\caption{ \it Contributions of the $O_{8g}$ operator to $B\to K \ell^+\ell^-$:
(a) factorizable, (b) and (c) nonfactorizable with soft  and hard gluon, respectively.}
\label{fig:O8}
\end{center}
\end{figure}

\begin{figure}[h]
\begin{center}
\hspace{-1 cm}
\includegraphics[scale=0.6]{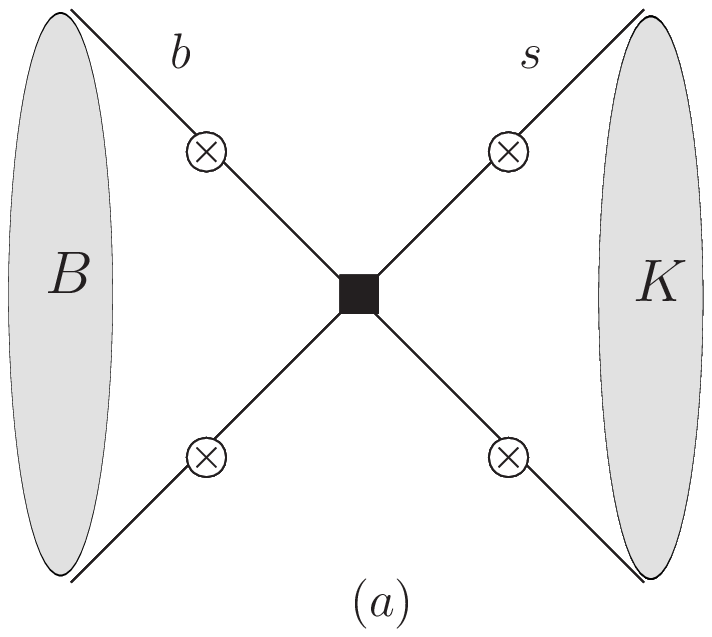}~~
\includegraphics[scale=0.6]{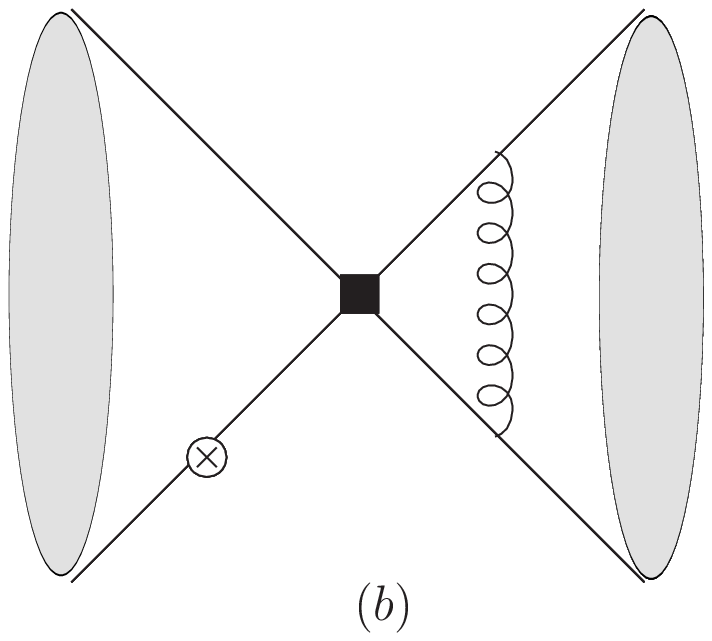}~~
\caption{\it Weak annihilation contribution to $B\to K \ell^+\ell^-$ amplitude: (a)
in LO and (b) one of the NLO  hard-gluon exchange diagrams. }
\label{fig:WA}
\end{center}
\end{figure}

The contributions to the nonlocal amplitude
(\ref{eq:matr}) are usually represented in a form of quark-gluon
diagrams shown one by one in Figs.\,\ref{fig:fact} - \ref{fig:WA}.
The dominant contribution
is generated by the  operators $O_{1,2}^{(c)}$  with large Wilson coefficients
and corresponds to
the diagrams shown in Figs.~\ref{fig:fact},\,\ref{fig:nonf}, where $q=c$  (the ``charm-loop'' effect).  This contribution includes
intermediate vector charmonium states  in the upper part of the decay kinematical region,
$m_{J/\psi}^2<q^2< (m_B-m_K)^2$.  Measuring
$B\to K\ell^+\ell^-$  in the vicinity of  $q^2=m^2_{\psi}$ ($\psi=J/\psi,\psi(2S), ...) $, one
practically observes a combination of the weak nonleptonic ($B\to \psi K$)
and leptonic ($\psi\to \ell^+\ell^-$)
decays, whereas the genuine FCNC  process $b\to s\ell^+\ell^-$, driven by the
effective operators $O_{7,9,10}$, turns into
a tiny background.  Note that the interval $4m_D^2< q^2< (m_B-m_K)^2$ also contains
broad charmonium resonances. To avoid the charmonium background, all experimental measurements
of $B\to K\ell^+\ell^-$  implement  a subtraction
of  the two $q^2$-bins around   $m^2_{J/\psi}$ and $m_{\psi(2S)}^2$.

In this paper we concentrate on the region of  small and intermediate
lepton-pair masses, $q^2<m_{J/\psi}^2$. In this region
the charm-loop effect is present in the
form of a virtual $\bar{c}c$ fluctuation. The latter
is usually approximated by $c$-quark loop diagrams,
starting from  the leading-order (LO)  simple loop  in Fig.~\ref{fig:fact}(a)
and including the  gluon  corrections.
A characteristic feature of this perturbative approximation at timelike $q^2$ is  a
``kink''  in the predicted differential width,  due to the onset of the imaginary
part of the $c$-quark loop at the threshold $q^2=4m_c^2<m_{J/\psi}^2$.
To avoid  this unphysical effect,
in what follows
we  employ the hadronic dispersion relation in the $q^2$ variable,
following \cite{KMPW}.
In this paper we also include the contributions
to ${\cal H} ^{(BK)}(q^2)$ stemming from the operators with light  quarks.
Hence, to stay away from quark-antiquark thresholds, all quark-gluon diagrams
shown in Figs.~\ref{fig:fact}-\ref{fig:WA} have to be calculated
at $q^2<0$ with sufficiently large $|q^2|$.

\begin{figure}[h]
\begin{center}
\hspace{-1 cm}
\includegraphics[scale=0.6]{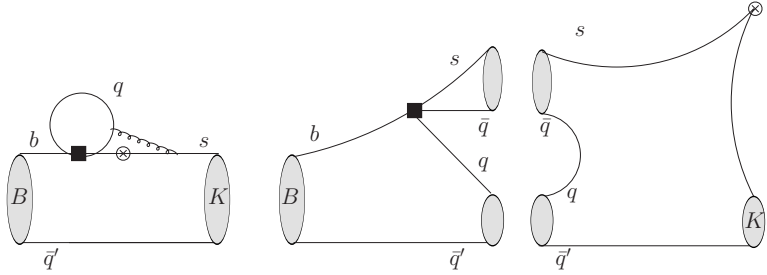}~~
\caption{\it  The NLO quark-loop diagram of $B\to K\ell\ell$ decay
(left) and one of its  hadronic  counterparts (right) containing  the intermediate pair of on-shell hadrons
with flavour content $(s\bar{q})(q\bar{q}')$.}
\label{fig:FSI}
\end{center}
\end{figure}

Before considering  separate contributions to the hadronic matrix element
${\cal H} ^{(BK)}(q^2)$, the following comment is in order.
A close inspection of the quark-loop diagrams at NLO shows that also at $q^2<0$,
below all hadronic thresholds in the e.m. current channel, the diagrams
develop an imaginary part.  This is a clear signal that
the intermediate hadronic states in the amplitude ${\cal H} ^{(BK)}(q^2<0)$
go on shell, so that the quark-gluon diagrams
provide only a local duality approximation for this amplitude.
For example, let us consider a diagram similar to the one
shown  in Fig.~\ref{fig:fact}(c), but with the
photon emitted from the $s$-quark. This diagram and one of  its
hadronic counterparts are shown in  Fig.~\ref{fig:FSI}.
To describe this hadronic transition in more detail  we choose $q=c,q'=d$
for definiteness.  In this case $\bar{B}^0$-meson decays to an on-shell
hadronic state with the quark content
$(\bar{c}s)(c\bar{d})$  (e.g., $\bar{B}^0\to D^-_s D^+$) which then
converts via strong and e.m. interaction into the final $\bar{K}^0$ meson and a
lepton pair
($D^-_s D^+\to \bar{K}^0 \ell^+\ell^-$).  This hadronic mechanism
involving multiple intermediate states
with the same quantum numbers
contributes to the strong final-state interaction in
$B\to  K \ell^+\ell^-$.
Approximating the sum over intermediate hadronic states
with a quark-level diagram is
justified by the large mass of the initial $B$-state and the large recoil
of the final hadron. This approximation
is similar to the one used in the QCD factorization  approach to
exclusive nonleptonic $B$-decays \cite{Beneke:2000ry}.

Let us emphasize another  important point.
The imaginary part  of the amplitude ${\cal H} ^{(BK)}(q^2)$ at spacelike $q^2$
is generated by the discontinuities in the variable
$(p+q)^2$. In $B\to K \ell^+\ell^- $
with an on-shell initial $B$ meson, this variable is fixed
at $(p+q)^2=m_B^2$.  To illustrate a generic two-variable kinematics,
one has to consider a correlation function where,
instead of the $B$-meson state in (\ref{eq:matr}),  a vacuum state is taken
and a quark current $\bar{q}^{\prime}\gamma_5b $ with $B_{q^\prime}$-meson quantum numbers
is added to the  time-ordered operator product,  with the
momentum squared $(p+q)^2$. The variable  $q^2$ is kept fixed and spacelike.
Still, the  correlation  function has discontinuities in the  variable $(p+q)^2$
located below the $B$-meson pole
and generated by intermediate  $(s\bar{q})(q\bar{q}')$ states.
Note that the presence of these ``parasitic'' intermediate states in the $B$-meson channel
of the correlation function is  caused by the absence of external 4-momentum
in the effective operator  vertex in  (\ref{eq:matr}).
Returning to the calculation of the hadronic amplitude ${\cal H} ^{(BK)}(q^2)$,
we conclude that  using OPE near the light-cone,
$x^2\sim 0$ in (\ref{eq:matr}) and restricting the calculation to $q^2<0$,
one cannot avoid using quark-gluon  diagrams at a timelike  (albeit large)
value of the second kinematical variable $(p+q)^2$.  Hence, strictly speaking,
the calculation procedure used here relies not entirely on OPE but also on
the local quark-hadron duality.

\section{Hadronic matrix elements in the spacelike region}
In this section we present
separate contributions to the hadronic amplitude ${\cal H}_{BK}(q^2<0) $
defined in (\ref{eq:matr}), including all effective operators entering $H_{eff}$ and taking
into account different possible quark topologies.

\subsection{Factorizable loop}

In LO, the contribution of the four-quark operators to $B\to K\ell^+\ell^-$
is shown in Fig.~\ref{fig:fact}(a).  In addition to the dominant $c$-quark loop
from the current-current operators
$O_{1,2}^{(c)}$ it also contains quark loops with various flavours,
originating from the quark-penguin operators $O_{3-6}$. In the adopted convention
for the effective operators the sum over all LO loops can be written as
\begin{eqnarray}
{\cal H}^{(BK)}_{fact,LO}(q^2) = {1 \over 8 \pi^2} \Bigg \{
Q_c\left(\frac{C_1}3+C_2\right)g(m_c^2,q^2) \nonumber
\\
+\left(\frac{C_3}3+C_4\right)\bigg[Q_s \, g(m_s^2,q^2)
+Q_b\,g(m_b^2,q^2)\bigg]
\nonumber \\
+\Bigg(C_3+\frac{C_4}3+
C_5+\frac{C_6}3\Bigg)
\sum\limits_{q=u,d,s,c,b}Q_q\,\bar{g}(m_q^2, q^2)
\Bigg \}f^+_{BK}(q^2)\,,
\label{eq:Afact}
\end{eqnarray}
where
\begin{eqnarray}
g(m_q^2,q^2) &=& -
\left (\ln {m_q^2 \over \mu^2} +1
\right ) + q^2    \int_{4 m_q^2}^{+\infty} ds   {
\sqrt{1- {4 m_q^2 \over s}} \left ( 1+ { 2 m_q^2 \over s} \right)
\over  s (s-q^2)} \,,
\label{eq:gfunct}
\end{eqnarray}
is the well-known loop function, $\mu$ is the renormalization scale
and $ \bar{g}(m_q^2,q^2) =
g(m_q^2,q^2)+1$. The definition of $g(m_q^2,q^2)$ used in \cite{KMPW}
differs from the above by the factor 4/9.
Since here we employ the loop diagrams at $q^2<0$,
that is below all quark-antiquark thresholds,  only the real part of
$g(m_q^2,q^2) $ is needed. Hereafter we also neglect the mass of the $u$ and $d$ quarks,
so that for them the loop function is  $g(0,q^2)=2/3-\ln(-q^2/\mu^2)$.

\subsection{Factorizable NLO contributions}
The NLO corrections to the quark loops
include the diagrams shown in Figs.~\ref{fig:fact}(b,c).
In the same order of perturbative expansion the gluon-penguin operator
enters with the diagrams in Fig.~\ref{fig:O8}(a).
The characteristic feature of all these diagrams \cite{BFS}  \footnote{
For an earlier estimate of this effect within quark model see, e.g.
\cite{Asatrian:1999mt}. }
 is that the corresponding hadronic matrix elements are  {\em factorizable}
in the same form as the LO contributions considered in the previous
section. Following \cite{BFS}, we extract the hard-scattering coefficient functions multiplying
 the $B\to K$ form factor from
the NLO diagrams for the $b\to s \ell^+\ell^-$
decay rates calculated in \cite{Asatrian:2001de,Asatryan:2001zw}.

Summing over all contributing operators, the factorizable NLO contributions
to the hadronic amplitude  can be written as
\begin{eqnarray}
{\cal H}^{(BK)}_{fact,NLO}(q^2) = - \frac{\alpha_s }{32 \pi^2}\frac{m_b}{m_B} \bigg \{ C_1 F_2^{(7)}(q^2) +  C_8^{eff} F_8^{(7)} (q^2)  \nonumber \\
+ {m_B \over 2 m_b} \bigg [C_1 F_2^{(9)}(q^2)  + 2 C_2 \left (
F_1^{(9)}(q^2)  + {1 \over 6} F_2^{(9)}(q^2)  \right ) + C_8^{eff} F_8^{(9)}(q^2)
\bigg ] \bigg \} f^+_{BK}(q^2) \,.
\label{eq:HfactNLO}
\end{eqnarray}
The definitions and nomenclature of the indices of
the functions $F_{1,2,8}^{(7,9)}$  are the same as in
\cite{Asatrian:2001de,Asatryan:2001zw}, where
$F_{1,2}^{(7,9)}$ are expressed as a double expansion in
$\hat{s}=q^2/m_b^2$ and $\hat{m}_c^2=m_c^2/m_b^2$.
It  has been shown in \cite{Asatrian:2001de,Asatryan:2001zw}
that   keeping the terms up to the third power of  $\hat{s}$  and
$\hat{m}_c^2$ provides a sufficient numerical accuracy
in the region  $0.05 \leq \hat{s}\leq 0.25 $.
Here we use these expansions
at $q^2<0$, restricting ourselves to $|\hat{s}|<0.25$.
For $F_{8}^{(7,9)}$ we use the expressions derived in \cite{BFS}.
Note that the NLO diagrams generate
an imaginary part in the coefficient functions entering  (\ref{eq:HfactNLO}), which, as discussed in the previous  section,
has to be interpreted as a quark-hadron duality
counterpart of the strong rescattering in the hadronic amplitude.

A dedicated  LCSR calculation
of the hadronic matrix element (\ref{eq:HfactNLO})  at $q^2<0$ is also possible. One has
to introduce a correlation function where the $B$ meson is represented
with its DA's and $K$ meson is interpolated with a quark current.
However, in the case of four-quark operators,
the complexity of the two-loop
diagrams with several momentum/mass scales emerging in  the OPE for the correlation
function makes the computation of hard-gluon contributions from LCSR's
a very difficult task.
On the other hand, the diagrams for the correlation function
with the gluon-penguin operator  have only one loop and can
in principle be calculated in future, to be compared
with the QCD factorization results which are used here.

\subsection{Nonfactorizable soft-gluon contributions}

The contributions to the hadronic amplitude ${\cal H}_{BK}(q^2) $
that are beyond QCD factorization, include also the soft-gluon emission from the quark loop,
as shown in Fig.~\ref{fig:nonf}(a). In \cite{KMPW} the dominant part of this effect,
caused by the $c$-quark loop
was estimated  using the light-cone OPE at $q^2\ll 4m_c^2$, where
this power correction is suppressed  by the inverse power of  $(4m_c^2-q^2)$ with respect
to the LO loop.
Here we also include the contributions of the soft-gluon emission from
the quark loops generated by the quark-penguin operators.
In presence of the light-quark loops, we shift the calculation to
$q^2<0$, so that $|q^2|\gg \Lambda_{QCD}^2$.

The $B\to K $  hadronic matrix element for the soft-gluon emission from the
loop with a generic quark flavour $q$ can be easily derived from the results of
\cite{KMPW}.  At $q^2\ll 4m_q^2$, the  soft-gluon emission
effect is reduced to a  hadronic matrix element
\begin{eqnarray}
\langle \bar{K}(p) | \tilde{\mathcal{O}}_{\mu} (m_q, q)|
B(p+q)\rangle =[ (p \cdot q) q_{\mu}-q^2p_{\mu} ]  \,
\tilde{\A}(m_q^2,q^2) \,,
\label{eq:Atild}
\end{eqnarray}
where the operator
$\tilde{\mathcal{O}}_{\mu} (m_q,q)$ is equal
to the nonlocal effective operator
$\tilde{\mathcal{O}}_{\mu} (q)$ defined in (3.14) of
\cite{KMPW}   with the  replacement $m_c \to m_q$ in the
coefficient function $I_{\mu \rho \alpha \beta}(q,\omega)$ given in (3.15).
Collecting the contributions from
all four-quark operators, we obtain for the soft-gluon emission  effect
\begin{eqnarray}
{\cal H}^{(BK)}_{soft,4q}(q^2) =
2 \big\{ Q_c (C_1+C_4-C_6) \tilde{\A}(m_c^2, q^2)
+(C_4-C_6) (Q_u+Q_d) \tilde{\A}(0, q^2)
\nonumber\\
+(C_3+C_4-C_6)[ Q_b \tilde{\A}(m_b^2, q^2) +
Q_s \tilde{\A}(m_s^2, q^2)]\big\}\,.
\label{eq:HBK}
\end{eqnarray}
Note that in this contribution the dominant charm-loop part has an enhanced
Wilson coefficient with respect to the one in the LO expression
(\ref{eq:Afact}).  At  $q^2<0$  the hadronic matrix element
$\tilde{\A}(m_q^2, q^2)$  is obtained from LCSR
with $B$-meson  DA's, using the analytic expression (4.8)
in \cite{KMPW}  derived for the $c$-quark case. The quark-flavour dependence is
concentrated in the denominator which stems from the coefficient
function of the effective operator.

Finally, there is also a  soft-gluon emission generated by the $O_{8g}$ operator
as shown in Fig. \ref{fig:O8}b. It is a nonfactorizable effect, so far not taken
into account. In the next section we will derive a dedicated LCSR for the
corresponding hadronic matrix element.

Note that within  the LCSR  approach a systematic separation of
``hard''- and ``soft''-gluon contributions, with high and low average
virtualities, respectively, is possible,
attributing them to different terms in the OPE.  E.g.,
in the sum rules with $B$-meson DA's  the soft-gluon
contributions  enter  the  terms with three-particle DA's,
whereas the hard-gluon contributions
enter the NLO coefficient functions  in the terms containing quark-antiquark DA's.
Here we  use LCSR's to compute the soft-gluon
contributions whereas  the hard-gluon effects
are approximated by QCD factorization.

\subsection{Nonfactorizable spectator contributions}

A highly-virtual (``hard'')  gluon emitted from the intermediate quark loop or
from the $O_{8g}$   operator vertex, can be
absorbed by the spectator quark in the $B\to K$ transition, generating
an additional  contribution -- see Fig.~{\ref{fig:nonf}(b) or
Fig.~{\ref{fig:O8}(c), respectively -- which cannot be reduced to the $B\to K$ form factors.
In \cite{BFS} a factorization formula for this contribution was derived
in terms of the hard-scattering kernels convoluted with the $B$-meson and kaon
DA's:
\begin{eqnarray}
{\cal H}^{(BK)}_{nonf, spect.}(q^2) =  {\alpha_s C_F \over 32 \pi N_c}
 { f_B f_K m_b \over m_B^2} \sum_{\pm} \int {d \omega
\over \omega} \phi_{B}^{\pm}(\omega) \int_0^1 d u \,\varphi_K(u)
T^{(1)}_\pm(u,\omega)\,,
\label{eq:nfspect}
\end{eqnarray}
where the standard definition (see e.g., \cite{Beneke:2000wa})
of two-particle $B$ meson DA's  $\phi_{B}^\pm(\omega)$ and
of the twist-2 DA of the kaon  $\varphi_K(u)$ is used.
The NLO hard kernels $T_{\pm}^{(1)}(u,\omega)$ corresponding
to the sum of the diagrams in Fig.~\ref{fig:nonf}(b) and Fig.~\ref{fig:O8}(c)
are  taken from  \cite{BFS}:
\begin{eqnarray}
T_{+}^{(1)}(u,\omega)&=& - {m_B \over m_b} \big [ Q_c
t_{||}(u,m_c)  (C_1 + C_4 - C_6 )
+ Q_b t_{||}(u, m_b) (C_3 + C_4 - C_6 ) \nonumber \\
&& + Q_s t_{||}(u, m_s) (C_3 + C_4 - C_6 )+ Q_u t_{||}(u, 0) (C_4-C_6) \nonumber \\
&&  + Q_d t_{||}(u, 0) (C_4-C_6) \big  ]\,, \label{eq:Tplus}
\end{eqnarray}
\begin{eqnarray}
T_{-}^{(1)}(u,\omega)&=& - Q_q { m_B \omega \over m_B \omega
-q^2 } \Big \{  { 8 m_B \over 3 m_b }  \big [ g(m_c^2,\bar{u} m_B^2 + uq^2)
(C_1+C_4+C_6) \nonumber \\
&& + g(m_b^2,\bar{u} m_B^2 + uq^2) (C_3+C_4+C_6)\nonumber \\
&& + g(0,\bar{u} m_B^2 + uq^2) (C_3+3 C_4+ 3C_6)  \nonumber \\
&& -{2\over 3}  (C_3 - C_5 -15 C_6 )\big ] +{8 C_8^{eff} \over  \bar{u}+uq^2/m_B^2}
\Big \} \,,
\label{eq:Tmin}
\end{eqnarray}
where $Q_q$ is the electric charge of spectator quark in the $B$
meson ($q=u,d$).
The function $t_{||}(u, m_q)$ convoluted with the $B$-meson DA $\phi_B^+$
arises from the two diagrams in Fig.~\ref{fig:nonf}(b) where the virtual
photon is emitted from the quark loop.   We recalculated and confirmed
the resulting expression  for $t_{||}(u, m_q)$  presented in  Eq.~(28) of \cite{BFS}.
In the remaining  nonfactorizable diagrams where the virtual photon is emitted
from the quarks of the initial or  final meson, only the DA
$\phi_B^-$ contributes. Furthermore, we compared the above expression with the dedicated
sum rule calculation. To this end, a correlation function
with $B$-meson DA's and kaon interpolating current was introduced
where the internal quark loop with the virtual photon emission was inserted.
The resulting LCSR reproduces the  QCD factorization expression (\ref{eq:nfspect})
if two conditions are satisfied: (1) the asymptotic kaon DA
$\varphi_K (u)= 6u(1-u)$  is substituted in (\ref{eq:nfspect})
and  (2) in LCSR the leading-order two-point QCD sum rule
for  the kaon decay constant is used
\footnote{ \,Similar
correspondence was found \cite{DeFazio:2005dx} between  the
LCSR for the $B \to  \pi$ form factor in soft-collinear effective theory
and the  factorization formula for the
hard-collinear spectator contribution to the same form factor.}.

The factorization formula (\ref{eq:nfspect})
by construction \cite{BFS}
only includes the leading power in the  heavy-quark  limit. We expect that LCSR
can be more advantageous in assessing the power-suppressed corrections to this  contribution. Note that in the factorization approach  these corrections  involve convolutions of the $B$-meson DA's with higher-twist kaon DAs and generally suffer from end-point divergences. In the  LCSR framework, albeit technically
more involved, the off-shell correlation
function is used where only the $B$-meson DAs enter.

\subsection{Weak annihilation}

For this contribution shown in Fig.~\ref{fig:WA}
we use the same approximation as in \cite{BFS}, employing
the LO  factorization formula
\begin{eqnarray}
{\cal H}^{(BK)}_{nonf,WA}(q^2) = {1 \over 8  N_c}
 { f_B f_K m_b \over m_B^2} \sum_{\pm} \int {d \omega
\over \omega} \phi_{B}^{\pm}(\omega) \int_0^1 d u \, \varphi_K(u)
T^{(0)}_{ \pm}(u,\omega)\,.
\label{eq:WA}
\end{eqnarray}
The hard-scattering  kernels $T_{ \pm}^{(0)}(u,\omega)$ are given by
\begin{eqnarray}
T_{ +}^{(0)}(u,\omega)&=& 0, \nonumber \\
T_{ -}^{(0)}(u,\omega)&=& e_q \, { m_B \omega \over m_B \omega
-q^2 } {4 m_B \over m_b} (C_3 + 3 C_4)\,.
\label{eq:TWA}
\end{eqnarray}

The leading-power contribution -- as  well known --
originates from  the virtual photon emitted from the
light-spectator quark inside the $B$-meson.  The other three diagrams
in Fig.~\ref{fig:WA}(a)  are essential to ensure
the e.m. current conservation.  As noticed in \cite{BFS}, the  weak  annihilation diagrams
develop an end-point singularity
when the invariant mass of lepton pair becomes soft,
$ q^2\sim \Lambda^2_{QCD} $. Hence working at large spacelike
$q^2$ we avoid this problem.  Radiative
corrections to the  weak  annihilation diagrams
(one of them shown in Fig.~\ref{fig:WA}(b)) involving hard gluons
are neglected since they have an additional $O(\alpha_s)$ suppression.
Note that a  soft-gluon emission accompanied by weak annihilation
apparently cannot be described by the above factorization formula
but instead can be studied using LCSR's.
Since there is yet additional power suppression involved
we will not dwell on this problem here, expecting that altogether
the above factorization formula provides a reasonable estimate
of this suppressed effect.

\subsection{ Power counting}

Concluding this section, let us discuss the power
counting of separate contributions to the
nonlocal amplitude ${\cal H} ^{(B K)}(q^2)$  considered  above.
As a reference scaling behaviour we take the $m_b\to \infty$  limit
of the dominant factorizable contributions to  $B\to K \ell^+\ell^- $ amplitude
(\ref{eq:ampl})
which are determined by the $B\to K$ form factors. Importantly, the large
spacelike scale  $|q^2|\gg \Lambda_{QCD}^2$
introduced in our calculation does not scale with $m_b$,
hence it does not influence the familiar scaling behavior
of the form factors in the large hadronic recoil region:
$f^+_{BK}\sim f^T_{BK}\sim 1/m_b^{3/2}$. This behaviour
follows from the power counting in LCSR's with $B$-meson DA's
for these form factors, as well as from the alternative LCSRs with
light meson (pion and kaon) DA's
(see e.g.,\cite{KMO2} for a detailed discussion). Note that in the
numerical calculation here we use the $B\to K$ form factors \cite{KMPW}
from the LCSR's with kaon DA's  calculated at finite $m_b$,
hence beyond the heavy-mass limit.

Turning to ${\cal H} ^{(BK)}(q^2)$,  we notice that the factorizable contributions
(\ref {eq:Afact}) and (\ref {eq:HfactNLO})  reveal the same
power counting $\sim 1/m_b^{3/2}$, since the coefficient functions  stemming from the loops can only introduce
a logarithmic dependence on the heavy mass scale. The scaling
of the nonfactorizable soft-gluon contribution (\ref{eq:HBK}) is less trivial;
it can be obtained from the heavy-quark limit of the
LCSR for the hadronic matrix element $\tilde{{\cal A}}(m_q^2,q^2) $
given in Eq.(4.8) in \cite{KMPW}.
The derivation of the heavy-quark limit follows the same procedure as for the
form factor LCSR's  in \cite{KMO2}.
We obtain
$\tilde{{\cal A}}(m_q^2,q^2)\sim \Lambda_{QCD}^2/(4m_q^2-q^2) \times O(m_b^{-3/2})$.
Furthermore, the power-counting  of the nonfactorizable spectator and
weak annihilation contributions is directly obtained from the factorization formulae
(\ref{eq:nfspect}) and (\ref{eq:WA}), respectively.
Taking into account the scaling behaviour  $f_B\sim 1/m_b^{1/2}$
and the fact that the kernels $g$ and $T_\pm^{(1,0)} $  in the leading power
are independent of $m_b$  (modulo logarithms) we obtain that
both these contributions scale as $O(m_b^{-3/2})$. Note however that numerically
they are quite different, the weak annihilation contribution (\ref{eq:WA})
being suppressed due to the Wilson coefficients.  Finally, the power-counting
analysis of the soft-gluon nonfactorizable contribution of $O_{ 8g }$
presented in the next section reveals an additional  $1/m_b^2$ suppression
with respect to the universal $1/m_b^{3/2}$ scaling behaviour.
Summarizing,  the nonlocal amplitude ${\cal H} ^{(B K)}(q^2)$
has the same leading-power behaviour
as the dominant contributions of the $O_{7,9,10}$ operators.
The presence of the additional spacelike scale $q^2$ is crucial
because the main nonperturbative
correction to this amplitude stemming  from the soft-gluon emission from the
quark loops, is ``protected'' by $\sim 1/|q^2|$ . The same $q^2$-scale
plays a role of a regulator in the nonfactorizable spectator and weak annihilation
contributions.  This important circumstance justifies our approach
of calculating the nonlocal amplitude at large spacelike $q^2$.
Future efforts to improve the adopted approximation,
e.g.,  in the framework of LCSR's, should be invested in calculating
the next-to-leading power corrections in $1/q^2$, such as the two soft-gluon emission
diagrams.

\section{Soft-gluon contribution of the $O_{8g}$ operator }

\begin{figure}[tb]
\begin{center}
\hspace{-1 cm}
\includegraphics[scale=0.5]{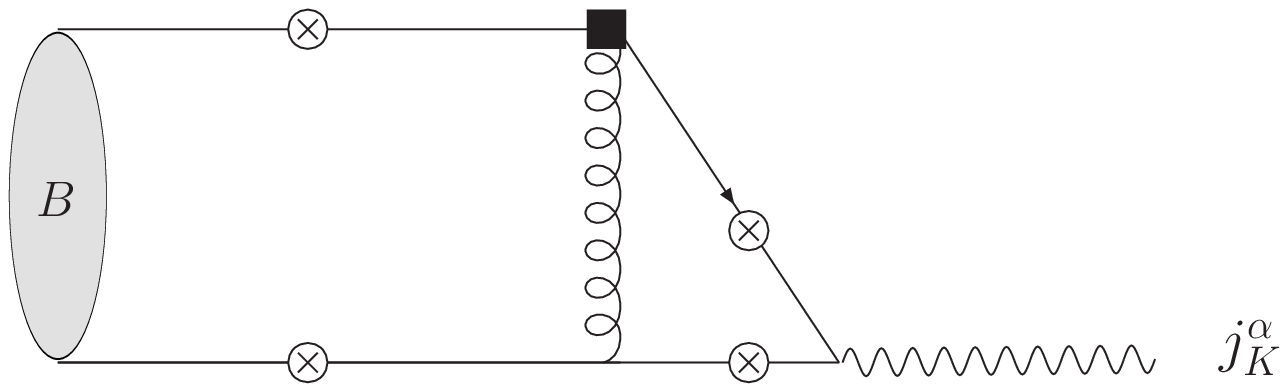}~~
\includegraphics[scale=0.5]{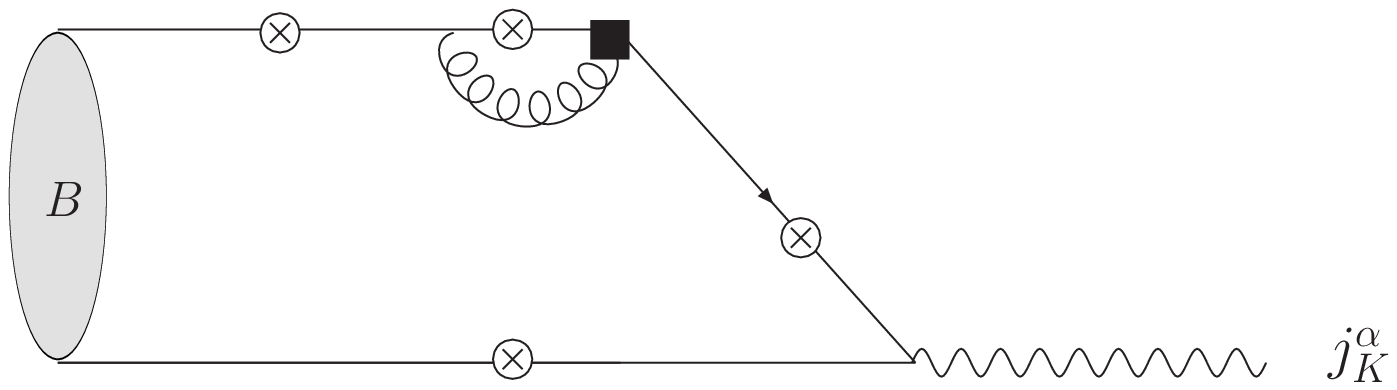}\\
(a) \hspace{5cm} (b)
\\[3mm]
\includegraphics[scale=0.5]{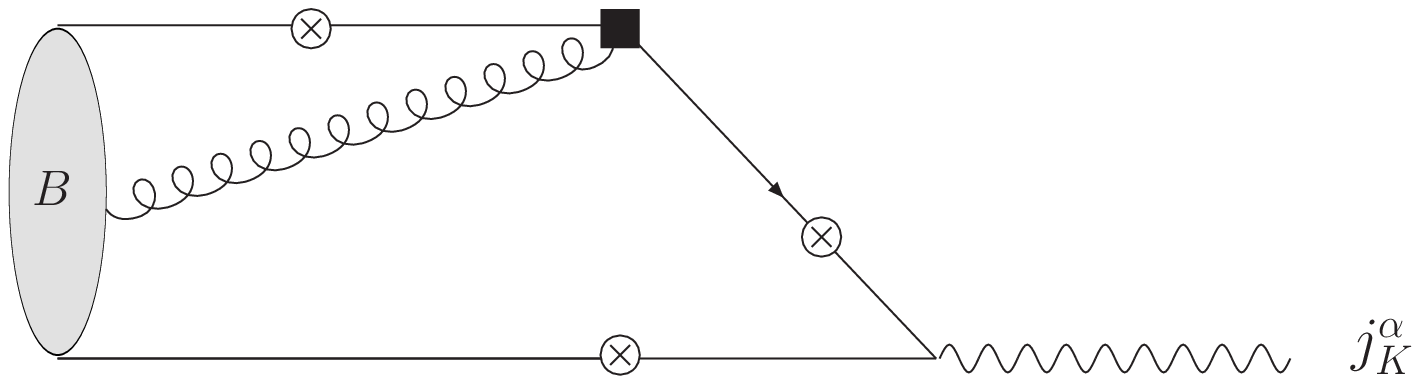}\\
(c)
\caption{\it Correlation function used to derive LCSR for the
gluon-penguin operator contribution. The diagrams (a),(b) and (c)
correspond to the hard-gluon and soft-gluon contributions, respectively.}
\label{fig:O8corr}
\end{center}
\end{figure}
Here we present a calculation
of the nonfactorizable effect  in $B\to K\ell^+\ell^-$
due to  the soft-gluon emission generated by the gluon-penguin operator $O_8$.
The diagram in Fig.\ref{fig:O8}(b) schematically represents this effect;
important is that the gluon emission is a long-distance effect
as opposed to diagrams Fig.\ref{fig:O8}(a,c) where the gluon is absorbed at short distances and can be described by a perturbative propagator.
The nonlocal hadronic matrix element
\begin{eqnarray}
i \int d^4x e^{q\cdot x}\langle K(p)| T \{j^{\mu}_{em}(x), C_{8}^{eff} O_{8g}(0) \}| B(p+q) \rangle =
[ (p \cdot q) q^{\mu}-q^2p^{\mu} ]  \, {\cal
H}^{(BK)}_{O_{8g}}(q^2)
\label{eq:HO8}
\end{eqnarray}
includes all effects depicted in Fig.~\ref{fig:O8}. As explained in the
previous section, the soft-gluon part of this amplitude
corresponds to the
term in LCSR with the quark-antiquark-gluon components of the $B$ meson
DA's. The derivation of the sum rule follows the procedure
used in \cite{KMPW}, and the operator $O_{8g}$  is
simpler than the effective nonlocal operator of the soft-gluon emission
from the quark loop.

We start from the following correlation function
\begin{eqnarray}
\Pi^{\alpha \mu}(p,q) = i \int d^4y \int d^4x  \, e^{i (p \cdot y
+ q \cdot x)}\langle 0 | T \{j_K^{\alpha}(y), j^{\mu}_{em}(x), C_{8}^{eff} O_{8g}(0)
 \}| B(p+q) \rangle  \nonumber \\
= i p^{\alpha} q^{\mu} \Pi(p^2,q^2) + \ldots \,,
\label{O8correlator}
\end{eqnarray}
where the gluon-penguin operator enters together with the e.m. current
and kaon interpolating current
$j_K^{\alpha}=\bar{d}\gamma^{\alpha} \gamma_5 s$.
It is sufficient to consider the single  kinematical structure
shown on r.h.s., the rest  is denoted  by ellipses.
At the hadronic level,  we can express the above correlation
function with the help of dispersion relation in the variable
$p^2$  (the kaon-current momentum squared) at fixed $q^2$:
\begin{eqnarray}
\Pi^{\alpha\mu}(p,q) = {i f_K p^{\alpha} \over m_K^2 -p^2} [(p\cdot q) q^{\mu}-q^2p^{\mu} ]  \, {\cal
H}^{(BK)}_{O_{8g}}(q^2)  + ...\,,
\label{eq:dispHO8}
\end{eqnarray}
where the ground-state term of the kaon contains the $B\to K$
transition amplitude we are interested  in,  and ellipses denote the contributions
from the excited and continuum  states with the kaon quantum numbers.
To derive the sum rule, we match the coefficient
at the kinematical structure $\sim p^{\alpha}q^{\mu}$ in the above
to the invariant amplitude $\Pi(p^2,q^2)$ in (\ref{O8correlator}).
At sufficiently large $|p^2|$ this correlation
function (\ref{O8correlator}) can be computed in terms of OPE
in terms of $B$-meson DA's. The gluon field
emitted from the operator $O_{8g}$ is either absorbed by one of the virtual quark lines,
or enters the $B$-meson state forming  three-particle DA's (see Fig.\,\ref{fig:O8corr}).
Only the latter term in OPE interests us here, hence we only have to
calculate the diagrams in Fig.~\ref{fig:O8corr}(c).
The diagrams in Figs.~\ref{fig:O8corr}(a,b) correspond to the hard-gluon contributions
already taken  into account in terms of QCD factorization
and included in the contributions ${\cal H}^{(BK)}_{fact,NLO}$
and ${\cal H}^{(BK)}_{nonf,spect}$,  as described  in the previous section.

We decompose the
three-particle Fock state of the $B$ meson in terms of
four DA's: $\Psi_A(\omega,\xi)$, $\Psi_V(\omega,\xi)$, $X_A (\omega,\xi)$ and $Y_A (\omega,\xi)$,
where $\omega$ and $\xi$ are the momenta carried by the
light-quark and gluon inside the $B$-meson. The relevant definitions and
ansatz for DA's (see App. B) are the same as the ones used in \cite{KMPW}.
After inserting this decomposition in the correlation function
we calculate the sum of two diagrams in Fig.~\ref{fig:O8corr}(c) with the virtual photon emission
from $b$ and $s$ quarks; note that the photon emission from the spectator
light quark  does not contribute to LCSR.

The invariant amplitude $\Pi(p^2,q^2)$
is obtained in the following form:
\begin{eqnarray}
\Pi(p^2,q^2)={ m_b \over 8 \pi^2 } f_B \sum_{n=1,2}
\int_0^{\infty} d \omega \int_0^{\infty} d\xi  \, { 1 \over
[(p-\omega v)^2-m_s^2 ]^n}   \nonumber \\
\times \bigg [ F_{n}^{(\Psi A)}(q^2,\omega) \Psi_A
(\omega,\xi) +  F_{n}^{(\Psi V)}(q^2,\omega) \Psi_V
(\omega,\xi)  \, &&
\nonumber \\
+ F_{n}^{(XA)}(q^2,\omega) \overline{X}_A (\omega,\xi) +
F_{n}^{(YA)}(q^2,\omega) \overline{Y}_A (\omega,\xi)  \bigg
]\,,
 \label{eq:QCDcorr}
\end{eqnarray}
where $\overline{X}_A (\omega,\xi)=\int\limits_0^\omega d\eta X_A (\eta,\xi)$,
$\overline{Y}_A (\omega,\xi)=\int\limits_0^\omega d\eta Y_A (\eta,\xi)$.
The explicit expressions of  the coefficient functions $F^{(DA)}_{n}(q^2,\omega)$
($n=1,\,2$)  multiplying the  three-particle $B$-meson DA's are collected in App.~C.
To simplify the calculation, we neglect the numerically very
small $\sim \xi/m_B$  terms in these functions,  hence in the leading power
they are independent of the gluon-momentum  variable $\xi$.

Equating two different representations of the correlation function,
we  obtain
\begin{eqnarray}
{f_K (m_B^2-m_K^2-q^2) \over 2 (m_K^2-p^2)} \,
{\cal H}^{(BK)}_{soft,O_8} + \int_{s_0^h}^{\infty}  ds \, {
\rho^{h}(s,q^2)\over s- p^2 } = \Pi(p^2,q^2) \,.
\end{eqnarray}
In the above, the index ``soft'' at the amplitude ${\cal H}^{(BK)} $ indicates
that we only take into account the soft-gluon contribution in the OPE result
for the correlation function.

The next important step is to use the quark-hadron duality
approximation for the
integral over the excited and continuum hadronic states with the kaon
quantum numbers:
\begin{eqnarray}
\int_{s_0^h}^{\infty}  ds \, { \rho^{h}(s,q^2)\over s- p^2 } = {1
\over \pi} \int_{s_0^K}^{\infty} ds { {\rm Im}\Pi(s,q^2) \over
s-p^2} \,,
\end{eqnarray}
where the effective threshold $s_0^K$  is introduced.
Finally, the Borel transformation is performed to improve the duality
approximation and the final sum rule reads:
\begin{eqnarray}
\, {\cal H}^{(BK)}_{soft,O_8}(q^2) = { 2 \, e^{m_K^2 / M^2}\over
f_K (m_B^2-m_K^2-q^2)} {1\over \pi} \int_{0}^{s_0^K} ds \, \,
e^{-s / M^2} \, {\rm Im}\Pi(s,q^2) \,.
\label{eq:SRO8}
\end{eqnarray}
To obtain the explicit form of the integral
on the r.h.s. it is convenient to start from the
initial expression (\ref{eq:QCDcorr}) and use the
substitution rules presented in App.~C for
separate integrals in this expression. These substitutions
simultaneously perform the transition to the
dispersion form, subtraction of the higher states and Borel
transformation.

The soft-gluon contribution of $O_{8g}$ is suppressed by an
extra power of $O(\Lambda^2/m_b^2)$ with respect to the
factorizable contributions, such as the LO loop contribution. Here $\Lambda$ is
a typical low-mass scale entering $B$-meson DA's in HQET.
The relevant power counting can be carried out by expanding the sum rule (\ref{eq:SRO8})
at $m_b\to \infty$.  Focussing, e.g., on the contribution of the
DA $\Psi_A(\omega,\xi)$   in  this sum rule, we obtain
after using the substitution rules in App.~C:
\begin{eqnarray}
\, {\cal H}^{(BK),(\Psi_A) }_{soft,O_8}(q^2)  &&\sim  {  m_b f_B \over
 (m_B^2-m_K^2-q^2)} \,
\int_0^{\omega_0} \frac{ d \omega \,  }{  1-\omega /m_B} e^{-s / M^2}
\nonumber \\
&&
\times \int_0^{\infty} d \xi  \, F_1^{(\Psi_A)}(q^2, \omega) \,  \, \Psi_A(\omega,\xi)
\sim   1/m_b^{7/2} \,,
\end{eqnarray}
taking into account that both the threshold parameter $\omega_0\sim s_0^K/m_B$ and the
coefficient function $ F_1^{(\Psi_A)}(q^2, \omega)$ scale as
$1/m_b$.

\section{Nonlocal hadronic effects in timelike region}

Collecting the results for separate contributions presented in the previous sections
we calculate the nonlocal hadronic amplitude
defined in (\ref{eq:matr}) at $q^2<0$:
\ba
{\cal H} ^{(BK)}(q^2) &=& {\cal H}^{(BK)}_{fact,LO}(q^2)+
{\cal H}^{(BK)}_{fact,NLO}(q^2)
+{\cal H}^{(BK)}_{nonf,spect}(q^2) \nonumber \\
&& + {\cal H}^{(BK)}_{soft,4q}(q^2)+
{\cal H}^{(BK)}_{soft,O_8}(q^2)+
{\cal H}^{(BK)}_{nonf,WA}(q^2)\,.
\label{eq:Htot}
\ea
To access the physical region $4m_\ell^2<q^2<(m_B-m_K)^2$   of the $B\to K\ell^+\ell^-$ decay,
following \cite{KMPW}, we use hadronic dispersion relation in the channel of virtual photon.
Here the following observation is important.
The amplitude  ${\cal H} ^{(BK)}(q^2)$ in  (\ref{eq:Htot})
can be decomposed in separate contributions
distinguished by the flavour of the quark
interacting with the virtual photon
via e.m. current. One simply
collects all terms in ${\cal H} ^{(BK)}(q^2)$   that are proportional to the
electric charge $Q_q$ of the quark $q$.
This flavour splitting allows one to establish a correspondence
between the part of  (\ref{eq:Htot})
with a given $Q_q$ and the hadronic states
with the quark content $\bar{q}q$ in the dispersion  relation.
For example, the part of the amplitude ${\cal H} ^{(BK)}(q^2)$
proportional to $Q_{u,d}$ ($Q_{s}$) is dual to the part of the
hadronic dispersion relation
containing  the vector mesons $\rho$ and $\omega$  ($\phi$)
and excited and continuum states with the same quantum numbers.
Here we neglect small mixing effects. Accordingly,
the part of the amplitude ${\cal H} ^{(BK)}(q^2)$  proportional  to  $Q_{c}$ generates
the vector charmonium states and open charm states in the timelike region.
Since $q^2<m_{\Upsilon}^2$ in $B\to K \ell^+\ell^-$, the part of the amplitude proportional to $Q_b$,
only  contributes to the nonresonance background in the dispersion relation.

We separate the contributions from different flavors of quarks in (\ref{eq:Htot})
in the following way:
\be
{\cal H} ^{(BK)}(q^2)= {\cal H} ^{(BK)}_{scb}(q^2)+ {\cal H} ^{(BK)}_{ud}(q^2)\,,
 \label{eq:Hdiv}
\ee
so that the first (second) term in the above  contains all contributions from
the different terms in (\ref{eq:Htot})  proportional to $Q_{s,c,b}$ ($Q_{u,d}$).
The reason why the $s,c,b$-quark contributions are not separated from each other
is that  in certain terms included in (\ref{eq:Htot})
e.g., in ${\cal H}^{(BK)}_{fact,NLO}(q^2)$,
the virtual photon emissions from $b$ and $s$ quarks  are interconnected
by gauge invariance. Also
the results of \cite{Asatryan:2001zw} used to calculate these contributions do not allow
one the separation of the $c$-quark component. In future,   a more detailed flavour
separation  can be achieved, performing a dedicated  analysis of the NLO two-loop diagrams.
Another important point is  that the photon emission
from the spectator light quarks also contributes to ${\cal H}^{(BK)}(q^2)$. Hence,
one has to specify the light flavour of  the decaying  $B$ meson. Our default choice
is $\bar{B}^0\to \bar{K}^0\ell^+\ell^-$.
To calculate the isospin asymmetry we will also consider $B^-\to K^-\ell^+\ell^-$.
In the approximation adopted in this paper, the small CP violation effects are
neglected, hence the same results are valid for $B^0$ and $B^+$ decays, respectively.

The separate hadronic dispersion  relations
for the two parts of the decomposition  (\ref{eq:Hdiv}) are:
\begin{eqnarray}
{\cal H} ^{(BK)}_{scb}(q^2)={\cal H} ^{(BK)}_{scb}(q_0^2)+
(q^2-q_0^2) \Big[ \sum_{V=\phi,J/\psi,\psi(2S)} \frac{\kappa_V
f_V |A_{B VK}| e^{i \varphi_V} }{(m_V^2-q_0^2)(m_V^2-q^2-
im_V\Gamma^{tot}_V)}
\nonumber\\
+\int_{s_0^h}^{\infty} ds
\frac{\rho_{scb}(s)}{(s-q_0^2)(s-q^2-i\epsilon)}\Big]\,.
\label{eq:disppsi}
\end{eqnarray}
and
\begin{eqnarray}
{\cal H} ^{(BK)}_{ud}(q^2)={\cal H} ^{(BK)}_{ud}(q_0^2)+
(q^2-q_0^2) \Big[ \sum_{V=\rho,\omega} \frac{\kappa_V f_V  |A_{B V
K}| e^{i \varphi_V} }{(m_V^2-q_0^2)(m_V^2-q^2-
im_V\Gamma^{tot}_V)}
\nonumber\\
+\int_{\tilde{s}_0^h}^{\infty} ds
\frac{\rho_{ud}(s)}{(s-q_0^2)(s-q^2-i\epsilon)}\Big]\,,
\label{eq:disprho}
\end{eqnarray}
 In the above, the decay constant of a vector meson $V$
with the polarization vector
$\epsilon_{V}$   is defined as:
$\langle 0 |j^\mu_{em}|V\rangle = \kappa_{V   }m_Vf_V\epsilon^{\mu}_{V}$;
the coefficients
$\kappa_{\rho}=1/\sqrt{2}$, $\kappa_{\omega}=1/ (3\sqrt{2})$,
$\kappa_{\phi}=-1/3$, $\kappa_{J/\psi}=\kappa_{\psi(2S)}=2/3$
 follow from the valence quark content of $V$.
As discussed in \cite{KMPW} one subtraction
at $q_0^2$ guarantees the convergence of the dispersion integrals.
We choose  $q_0^2=-1.0$ GeV$^2$ and use the calculated result to fix
the subtraction terms ${\cal H}_{scb,ud} ^{(BK)}(q_0^2)$.
The advantage of  the flavour separation done above is that the
suppressed part of the nonlocal amplitude,
${\cal H}_{ud} ^{(BK)}(q^2<0)$, has its own dispersion
relation, allowing one a more accurate estimate
 of  contributions  with $\rho, \omega$ quantum numbers
in the timelike region, which otherwise would have been hidden
under the dominant contributions of charmonium states.

The residues of vector meson poles in both dispersion relations
(\ref{eq:disppsi}) and (\ref{eq:disprho})
contain amplitudes $A_{BVK}$ of the nonleptonic decays $B\to VK$.
Final-state strong rescattering is taken into account by the phases attributed to each amplitude
\footnote{ As discussed above, these phases are related to the discontinuities
in the second variable $(p+q)^2$. In this sense the dispersion
relations we are considering here can be interpreted as a
double-dispersion representations for the amplitude ${\cal H} ^{(BK)}((p+q)^2,q^2)$,
where each amplitude $A_{B V
K}$  is itself expressable in a form of single dispersion relation over $(p+q)^2$.}.
The $B\to V K$ decays are well measured, allowing one
to extract the absolute values of their amplitudes. The relative phases
of these amplitudes are unknown
and they may considerably influence the pattern of vector meson contributions
in the dispersion relations.
Note that  the absolute values $|A_{BVK}|$ extracted from the experiment
deviate from their naive factorization estimates obtained multiplying  $f_V$
with the combination of Wilson coefficients and the $B\to K$ form factor.
These deviations are considerable for $V=J/\psi,\psi(2S)$,  revealing large
nonfactorizable effects in $B\to J/\psi K, \psi(2S) K $, they are somewhat smaller
for the decays into light vector mesons. Altogether,
the dispersion relations (\ref{eq:disppsi}) and (\ref{eq:disprho})
contain too many hadronic degrees of freedom to be fixed and/or constrained
solely from the matching of the r.h.s. with the result of the calculation at negative $q^2$.
Therefore,
following the approach advocated in \cite{KMPW} we fix the
absolute values of residues
in both dispersion relations,  employing experimental
data on the decay constants and nonleptonic
decay amplitudes.  The main difference between the relations considered
in \cite{KMPW} and here is that, in addition to the contributions of charmonium
states, we also include the contributions of light vector mesons
employing the data on $B\to \phi K$ and $B\to \rho (\omega) K$ partial
widths.  Importantly, the data currently accumulated in \cite{PDG,Amhis:2012bh}
allow one to fix the decay amplitudes for both $\bar{B}^0$ and $B^-$
mesons separately, which is particularly important for the evaluation of the
isospin asymmetry.

After fixing the residues of the lowest resonance poles in both dispersion relations
 (\ref{eq:disppsi}) and (\ref{eq:disprho}), they still  contain  a significant amount
of unknown hadronic degrees of freedom from higher states, accumulated in
the integrals over spectral densities. Note that the
integral in (\ref{eq:disppsi}) contains the contributions of excited
$\phi$-resonances and continuum states with the same quantum numbers,
as well as
the contributions of all vector charmonium states and open charm-anticharm
states above the  threshold $s=4m_D^2$.  The integral in (\ref{eq:disprho})
in its turn, accumulates excited and continuum states with
the $\rho,\omega$ quantum numbers. At very large $s$, approaching the
$B_s^*$ meson mass squared the $\bar{b}s$ states contribute in both
integrals. At low and intermediate $q^2$ the presence
of these states is signaled (at  least in the factorizable part of
${\cal H} ^{(BK)}(q^2)$) by the growth of the $B\to K$ form factor.
The fit of the dispersion relation to the calculation result
at negative $q^2$ will introduce  important constraints
on the hadronic integrals entering  dispersion relations.
However, in order to perform a continuation to positive $q^2$
one needs to parametrize these integrals in terms of
a certain ansatz.

Let us first concentrate on the dominant part of the amplitude
parametrized by the hadronic dispersion relation (\ref{eq:disppsi}).
Note that the region of $q^2$ we are interested in, is located below
the open charm threshold $4m_D^2$.  The only (subdominant)
contribution in this region to the spectral density $\rho_{scb}(s)$
originates from the superposition of $\bar{s}s$ states
starting from $\phi$ meson  and including its excitations
and multiparticle states, such as $\bar{K}K$ with $J^P=1^-$.
The ansatz used in QCD sum rules \cite{SVZ} to describe
the $\bar{s}s$ -channel with vector quantum numbers is  the $\phi$ -meson pole
combined with  the continuum, the latter approximated by the quark-parton duality
with an effective threshold. This motivates us to
replace the spectral density by the following ansatz:
\begin{equation}
\rho_{scb}(s)\theta(s-s_0^h)= \frac{1}{\pi}
\mbox{Im} {\cal H}^{BK}_{scb}(s)\theta(s-s_0^\phi)\,,
\label{eq:phiansatz}
\end{equation}
where the effective threshold $s_0^\phi$ is determined from
the QCD sum rule in the $\phi$ channel.
The simplest choice  for the r.h.s. is to approximate it with the LO factorizable part
${\cal H}^{BK}_{fact,LO,s}$  where the index $s$ indicates
that only the part proportional to $Q_s$ is taken in (\ref{eq:Afact}).
The spectral density is  then reduced to the imaginary part of the
$s$-quark loop function (\ref{eq:gfunct}) multiplied  by the $B\to K$ form factor.
Note that in this case  we adopt a stronger assumption of
local duality, rather than the semi-local duality used in QCD sum rules.
Also at  $s< 4m_D^2$, that is within the validity region for LCSR,
the calculated form factor can still be used.
We therefore subdivide the integral in the dispersion relation (\ref{eq:disppsi})
in two parts and adopt the approximation (\ref{eq:phiansatz}) for the
interval $s_0^\phi<s<4m_D^2$.
In addition, the imaginary part of the integral at $q^2>s_0^\phi$ is modified
by introducing an effective width
factor in the denominator, replacing
$s-q^2-i\epsilon \to s-q^2-i\sqrt{s} \,\, \Gamma_{eff}(s)$. The dependence
of this width factor on $s$ is taken as $\Gamma_{eff}(s)=\gamma \sqrt{s}\Theta(s-4m_K^2)$,
following the resonance model adopted  for the timelike form factors
in \cite{BKK} where the universal value $\gamma=0.2$ was derived from the spectrum
of light vector mesons. The step-function accounts for the kinematical limit where
the width vanishes.

In the remaining integral from   $4m_D^2$ to
infinity,  the spectral density  $\rho_{scb}(s)$ cannot be simply parametrized
because it contains a complicated  superposition of four hadronic components:
1) broad charmonium resonances
and open-charm continuum states  (the dominant contribution),
2) the ``tail'' of $\bar{s}s$ states,
3) the states with $B_s^*$ quantum numbers and
4) the $\bar{b}b$ states corresponding to the photon emission from $b$ quark.
Note however, that in the region of our interest $q^2<4m_D^2$
(practically, below the $J/\psi$ mass),
this integral has no singularities and can be  represented as a generic series expansion
in powers of $q^2/4m_D^2$. An alternative is to perform a usual $q^2\to z$ transformation
and to use the $z$-expansion.
As a default model we use a rather simple approximation
\begin{eqnarray}
\int_{4m_D^2}^{\infty} ds
\frac{\rho_{scb} (s)}{(s-q_0^2)(s-q^2-i\epsilon)} &=& a_{scb} + b_{scb} \,  {q^2
\over 4 m_D^2}\,,
\label{eq:model_larges}
\end{eqnarray}
where $a_{scb}$  and $b_{scb}$ are unknown complex constants.

Turning to the  second dispersion relation (\ref{eq:disprho}),
we use a similar approximation for the spectral density:
\begin{equation}
\rho_{ud}(s)\theta(s-\tilde{s}_0^h)= \frac{1}{\pi}
\mbox{Im} {\cal H}^{BK}_{ud}(s)\theta(s-s_0^\rho)\,,
\label{eq:rhoansatz}
\end{equation}
where $s^\rho_0$ is the effective threshold, typical for the QCD sum rules
in the $\rho$ channel. Accordingly, the integral over large $s$
is represented similar to (\ref{eq:model_larges}) with the two
additional
complex parameters $a_{ud}$ and  $b_{ud}$.  For the sake of uniformity
we put the lower limit of this integral to the same value $4m_D^2$.

After parametrizing the integrals over the spectral densities,
the dispersion relations  (\ref{eq:disppsi}) and (\ref{eq:disprho}) are fitted
to the results for ${\cal H}^{BK}_{scb}(q^2)$ and  ${\cal H}^{BK}_{ud}(q^2)$
respectively, at negative $q^2$. The fitted
parameters are the coefficients $a_{scb},b_{scb}$
in the  expansion  of the integral (\ref{eq:model_larges})
and $a_{ud}$, $b_{ud}$ for the corresponding integral  for $u,d$ quark states.
After that, the two dispersion relations are added together and continued to
timelike $q^2$ resulting in  the desired nonlocal amplitude ${\cal H}^{BK}(q^2)$
for $B\to K \ell^+\ell^-$ decay.
To assess the dependence on the parametrization
of the dispersion relations,  apart from the default parametrization
of the dispersion integrals described above,
we  also considered three alternative versions:
(I) the integral (\ref{eq:model_larges})  is replaced with an effective pole
as it was done in \cite{KMPW};  (II)  a generic $z$-series parametrization
for the integral over higher states
(\ref{eq:model_larges}) is used, and (III) as an extreme choice
only the dominant $J/\psi$ and $\psi(2S)$ states
are left in the dispersion relation with no flavour-splitting applied.

\section{Numerical results}
\begin{figure}[tb]
\begin{center}
\hspace{-1 cm}
\includegraphics[scale=0.6]{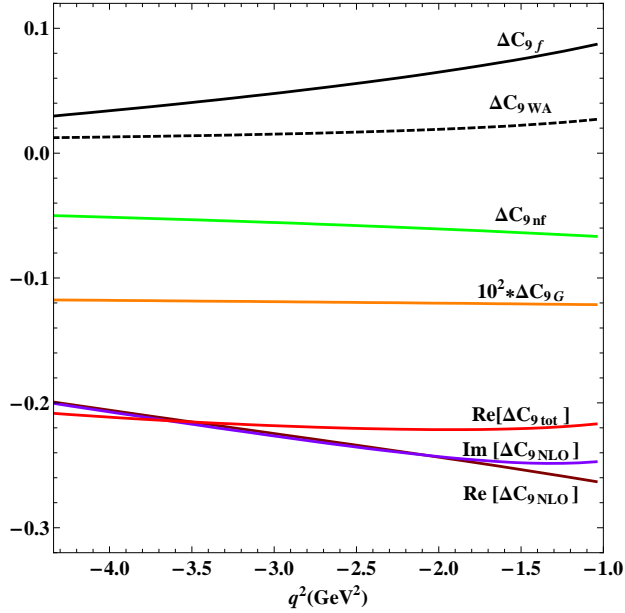}
\caption{\it Nonlocal hadronic amplitude in $B\to K\ell^+\ell^-$, expressed in terms of
the correction to $C_9$,  calculated at $q^2<0$
for the central values of the input. The separate contributions
are from the factorizable quark loops ($\Delta C_{9f}$, black),
 weak annihilation ($\Delta C_{9WA}$, dashed black),
 soft-gluon nonfactorizable emission
($\Delta C_{9nf}$, green),   soft-gluon emission from $O_{8g}$ operator
($\Delta C_{9G}$, multiplied by 10\,$^2$, orange), real part of NLO hard-gluon effects
($Re[\Delta C_{9\,NLO}]$, brown), total real part ($Re[\Delta C_{9 \,tot}]$, red),
imaginary part of  NLO hard-gluon effects ($Im[\Delta C_{9\,NLO}]$, violet).}
\label{fig:deltaC9}
\end{center}
\end{figure}

We start  with the numerical evaluation of the amplitude
 (\ref{eq:Htot}) in the region $q^2<0$.
The same input as in \cite{KMPW} is adopted, where one can find more details.
In particular, since the calculation is done at spacelike momentum transfer,
it is more appropriate to use the $\overline{ \rm MS}$ quark masses, for which
we adopt the following intervals:
 $m_s (2 \rm~GeV)= (98 \pm 16) ~\rm MeV$,  $m_c(m_c)=(1.29 \pm 0.03)~\rm GeV $
and   $m_b(m_b)=(4.164 \pm 0.025)~ \rm GeV$.
The functional form of $B$-meson DA's entering the factorization formulae  and LCSR's
is specified in App.~B.  We use the interval
$f_B=180\pm 30 $ MeV for the decay constant,
$\lambda_{B}(1 \rm~GeV) =460 \pm 110 ~\rm MeV$
for the inverse moment
and $\lambda_E(1 \rm~GeV)= \sqrt{3/2} \, \lambda_B ( 1 \rm~GeV)$
for the parameter entering the 3-particle DA's. For the kaon
decay constant  we use \cite{PDG} $f_K=159.8 \pm 1.4 \pm 0.44 \rm ~MeV $
and for the Gegenbauer moments of the kaon DA:
$a_1^K(1 \rm~GeV)=0.10 \pm 0.04$, $a_2^K(1 \rm~GeV)=0.25 \pm 0.15$.
Finally, in LCSR with the kaon interpolating current the
Borel parameter and effective duality threshold are taken
as $M^2= 1.0 \pm 0.25 \rm ~GeV^2$ and   $s_0^K=1.05 \rm ~GeV^2$, respectively.
The default value of the renormalization scale is $\mu=3.0 $ GeV,
so that $\alpha_s(3 \rm~GeV)= 0.252$, and the scale is then varied within
$2.5 \rm~GeV <\mu <4.5 \rm~GeV$.

For the $B\to K$ form factors the predictions from
LCSR with kaon DA's are used. They are presented
in the App.~B of \cite{KMPW}, in particular,
the normalization of the vector and tensor
form factors are given by $f^+_{BK}(0)=0.34^{+0.05}_{-0.02}$,
and  $f^T_{BK}(0)=0.39^{+0.05}_{-0.03}$, respectively.  In the contributions calculated
using QCD factorization approach we adopt a universal $B \to
K$  form factor $\xi_{BK}(q^2)$ which is set equal to $f^+_{BK}(q^2)$,
neglecting the small difference between tensor and  vector form factors.

According to (\ref{eq:deltc9}),
the result for ${\cal H}^{(BK)}(q^2<0)$ is normalized to the form factor,
yielding a correction to
the Wilson coefficient $C_9$. Separate contributions
to  this correction  are plotted  in Fig.~\ref{fig:deltaC9} for the central values of the input.
Numerically,  in the real part of $\Delta C^{(BK)}_9(q^2)$ there are substantial
cancellations between separate contributions, and the soft-gluon nonfactorizable
part plays an important role as already observed  in  \cite{KMPW}}.
Altogether,  the real and imaginary parts of $\Delta C^{(BK)}_9(q^2<0)$
-- the latter generated  by the NLO hard-gluon effects --
reach  the level of  a few percent of the short-distance coefficient $C_9$.

To assess the impact of nonlocal hadronic effects on the observables in $B\to K\ell^+\ell^-$,
we turn now to the numerical analysis of the
dispersion relations, allowing us to smoothly continue these effects
into the physical region $q^2>0$.
To this end, we follow the procedure described in the previous section,
splitting ${\cal H}^{(BK)}(q^2) $  according to (\ref{eq:Hdiv}),
and fitting the functions  ${\cal H}^{(BK)}_{scb}(q^2) $ and ${\cal H}^{(BK)}_{ud}(q^2) $
in the region  $-4 m_c^2 <q^2<-1.0$ GeV$^2$ to
the two dispersion relations (\ref{eq:disppsi}) and
 (\ref{eq:disprho}), respectively. In these relations,
the absolute values of the nonleptonic $B\to V K$ amplitudes, together with the
decay constants of vector mesons entering the residues
of the resonance poles, are extracted from
the experimental data \cite{PDG} on  the $B\to V K$  and $V\to \ell^+\ell^-$
widths.
These parameters  are collected in  Table \ref{tab:BVK}.
\begin{table}[h]
\begin{center}
\begin{tabular}{|c|c|c|c|c|c|}
\hline
Vector meson & $\rho$ & $\omega$ & $\phi$ & $J/\psi$ & $\psi(2S)$\\
\hline
&&&&&\\[-3mm]
$f_{V} \,\,$ & $221^{+1}_{-1}$ & $195^{+3}_{-4}$ & $228^{+2}_{-2}$ & $416^{+5}_{-6}$ & $297^{+3}_{-2}$ \\[2mm]
\hline
&&&&&\\[-3mm]
$|A_{\bar{B}^0  V \bar{K}^0}| $ & $1.3^{+0.1}_{-0.1}$
& $1.4^{+0.1}_{-0.1}$ & $1.8^{+0.1}_{-0.1}$& $33.9^{+0.7}_{-0.7}$ & $44.4^{+2.2}_{-2.2}$ \\[2mm]
&&&&&\\[-3mm]
$|A_{B^-  V K^-}| $ & $1.2^{+0.1}_{-0.1}$ &
$1.5^{+0.1}_{-0.1}$  &  $1.8^{+0.1}_{-0.1} $& $35.6^{+0.6}_{-0.6}$ & $42.0^{+1.2}_{-1.2}$ \\[2mm]
\hline \hline
\end{tabular}
\caption{ \it Decay constants of vector mesons
and amplitudes of $B\to V K$ decays
(all in MeV) calculated from the experimental data \cite{PDG}. }
\label{tab:BVK}
\end{center}
\end{table}

Furthermore, the integrals over  the spectral functions of higher states in (\ref{eq:disppsi}) and
 (\ref{eq:disprho}) are
subdivided in two parts. The integral below $4m_D^2$
is parametrized employing the local duality approximation
with the effective thresholds $s_0^\phi=1.95 $ GeV$^2$  in (\ref{eq:phiansatz})
and  $s_0^\rho=1.5 $ GeV$^2$  in (\ref{eq:rhoansatz}),
whereas for the remaining integrals above $4m_D^2$  the polynomial approximations
in the form (\ref{eq:model_larges}) are used.  In  (\ref{eq:disppsi}) the phases  of $J/\psi$ and
$\psi(2S)$  contributions are varied independently at
$- \pi <\varphi_{J/\psi}, \varphi_{\psi(2S)} \leq \pi$ and  the
$\phi$ contribution  is taken real.
For each combination of the phases $\varphi_{J/\psi}$ and
$\varphi_{\psi(2S)}$ the fit of  the complex parameters $a_{scb}$ and $b_{scb}$ is repeated.
The best fit  is obtained when a destructive interference
between the $J/\psi$  and $\psi(2S)$  terms in (\ref{eq:disppsi})
takes place,  as observed before in \cite{KMPW}.
The second dispersion relation (\ref{eq:disprho}) is treated in a similar way.
\TABLE[b]{
\begin{tabular}{|c|c|c|c|c|c|}
  \hline
  $\varphi_{J/\psi}$ & $\varphi_{\psi(2S)}$ & $|a_{scb}|$ [GeV$^{-2}$]&
${\rm Arg}[a_{scb}]$  & $|b_{scb}|$[GeV$^{-2}$] & ${\rm Arg}[b_{scb}]$  \\
  \hline
  $ -2.14 $ & $0.77$  & $1.40 \times 10^{-5}$ & $0.69$ & $1.37 \times 10^{-4}$ &
$0.86$ \\
  \hline
  \hline
  $\varphi_{\rho}$ & $\varphi_{\omega}$ & $|a_{ud}|$ [GeV$^{-2}$]& ${\rm Arg}[a_{ud}]$ &
$|b_{ud}|$ [GeV$^{-2}$]& ${\rm Arg}[b_{ud}]$  \\
  \hline
  $0.64$ & $-2.50$ & $2.74 \times 10^{-5}$ & $-2.13$ & $5.75 \times 10^{-5}$ &
$-2.07$ \\
  \hline
\end{tabular}
\caption{Parameters of the dispersion
relations (\ref{eq:disppsi}) and (\ref{eq:disprho})  obtained from the fit
to the amplitudes ${\cal H}^{(BK)}_{scb}$ and ${\cal H}^{(BK)}_{ud}$ at $q^2<0$
for the central values of the input. }
\label{table:fit}}
The phases and fitted parameters are collected in  Table \ref{table:fit}.
Substituting  them in the dispersion relations,
and continuing the latter to $q^2>0$ we finally  obtain a
numerical result for the  nonlocal hadronic amplitude ${\cal H} ^{BK}(q^2)$
in the physical region.
\begin{figure}[t]
\begin{center}
\hspace{-1 cm}
\includegraphics[scale=0.6]{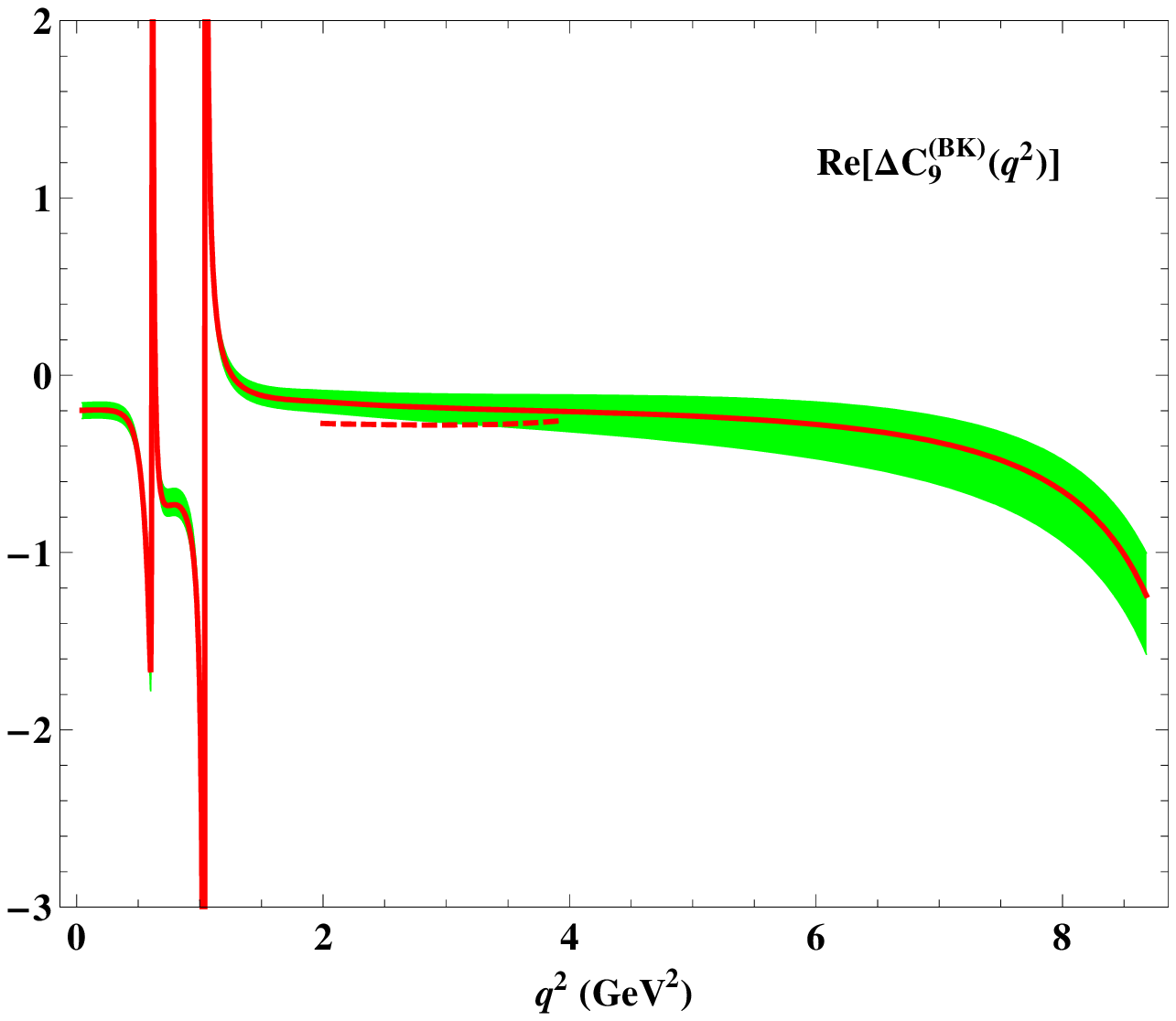}
\includegraphics[scale=0.6]{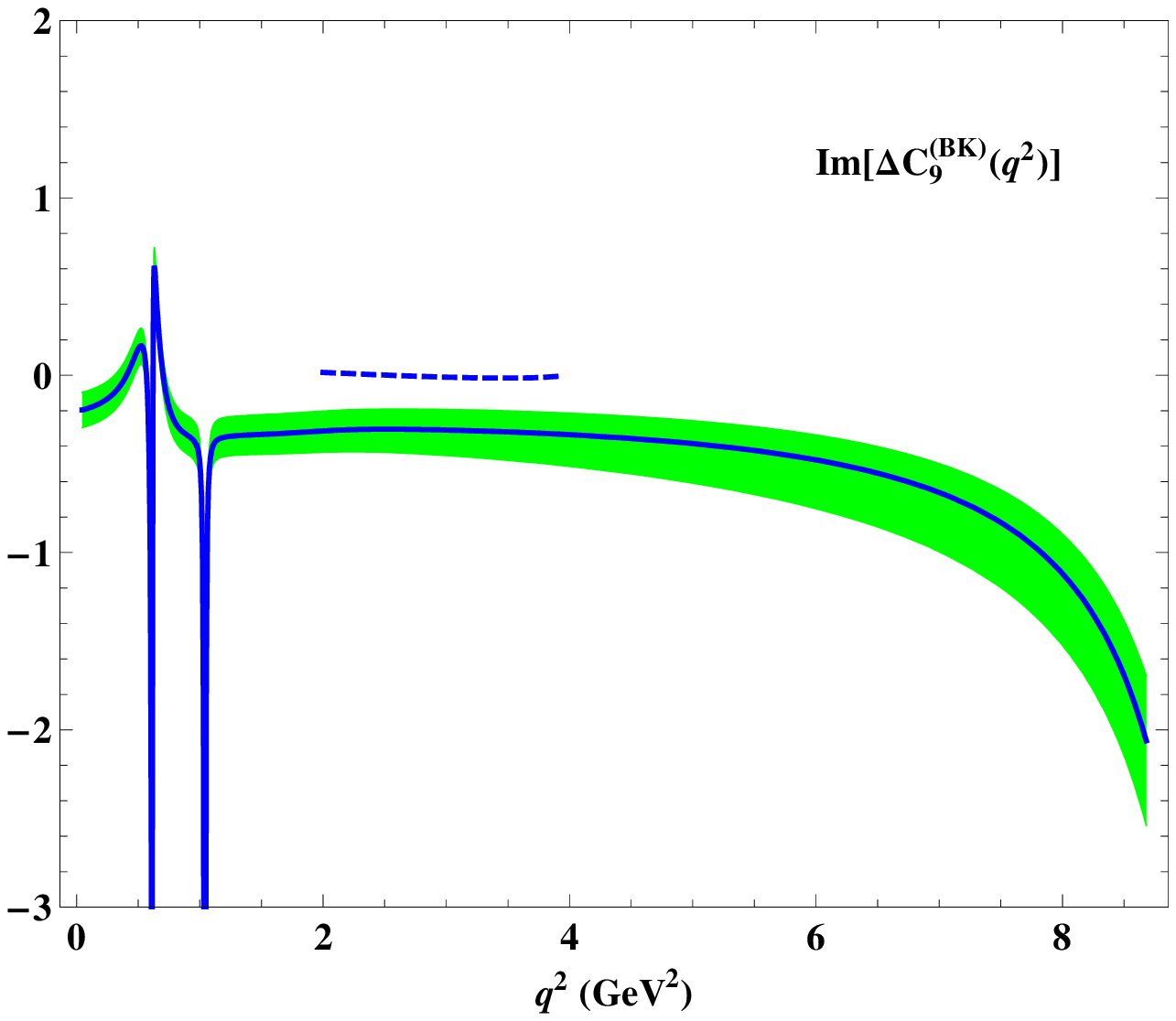}
\caption{\it
The effective correction $\Delta C^{(BK)}_9(q^2)$
in the physical region of $B\to K \ell^+\ell^-$: the red (blue) solid
 curve corresponds to the real (imaginary) part obtained from the
 hadronic dispersion relation, fitted to the QCD calculation at $q^2<0$
(central input, default parametrization).  The shaded areas indicate the uncertainties.
The dashed curves correspond to the  prediction of QCD factorization obtained with
same input.}
\label{fig:timl}
\end{center}
\end{figure}

The resulting effective correction $\Delta C_9^{BK}(q^2>0)$ is plotted
in Fig.~\ref{fig:timl}  in the region of  large hadronic recoil, up to
$q^2\sim m_{J/\psi}^2$. At $q^2 \leq m_\phi^2$  the behavior of
$\Delta C_9^{BK}(q^2)$ reflects the presence
of light vector resonances.
In the same figure we compare our result
with the result for  $\Delta C_9^{BK}$
directly calculated at   $q^2>0$, using the same approach
and approximation as in \cite{BFS}.
This comparison is possible in a restricted region,
$2.0\, {\rm GeV}^2\leq q^2\leq 4.0\, {\rm GeV}^2$,
above the light resonances and sufficiently below the $\bar{c}c$-quark  threshold.
In this region, the real parts $\Delta C_9^{BK}(q^2)$ obtained from
our dispersion approach and from the direct calculation are
in agreement within uncertainties, whereas the imaginary parts deviate
from each other.
\begin{table}[t]
 \caption{The effective correction to the coefficient $C_9$  at
 different $q^2$. }
\label{tab:res}
 \begin{center}
 \begin{tabular}{|c|c|c|}
   \hline
   $q^2$ (${\rm GeV}^2$) & ${\rm Re}[\Delta C_9^{BK}(q^2)]$ & ${\rm
 Im}[\Delta C_9^{BK}(q^2)]$  \\
   \hline
   $1.5$ & $-0.11^{+0.06}_{-0.05}$ & $-0.33^{+0.11}_{-0.11}$ \\
   \hline
   $2.0$ & $-0.15^{+0.07}_{-0.06}$ & $-0.31^{+0.11}_{-0.12}$ \\
   \hline
   $2.5$ & $-0.17^{+0.07}_{-0.07}$ & $-0.30^{+0.11}_{-0.13}$ \\
   \hline
   $3.0$ & $-0.18^{+0.08}_{-0.08}$ & $-0.31^{+0.12}_{-0.14}$ \\
   \hline
   $3.5$ & $-0.20^{+0.09}_{-0.10}$ & $-0.32^{+0.12}_{-0.16}$ \\
   \hline
   $4.0$ & $-0.21^{+0.09}_{-0.12}$ & $-0.34^{+0.12}_{-0.18}$ \\
   \hline
   $4.5$ & $-0.22^{+0.10}_{-0.13}$ & $-0.36^{+0.13}_{-0.20}$ \\
   \hline
   $5.0$ & $-0.23^{+0.11}_{-0.15}$ & $-0.39^{+0.13}_{-0.22}$ \\
   \hline
$5.5$ & $-0.25^{+0.12}_{-0.17}$ & $-0.42^{+0.13}_{-0.25}$ \\
   \hline
   $6.0$ & $-0.28^{+0.13}_{-0.19}$ & $-0.48^{+0.14}_{-0.28}$ \\
   \hline
   $6.5$ & $-0.32^{+0.14}_{-0.22}$ & $-0.55^{+0.15}_{-0.30}$ \\
   \hline
   $7.0$ & $-0.38^{+0.15}_{-0.24}$ & $-0.66^{+0.16}_{-0.33}$ \\
   \hline
   $7.5$ & $-0.48^{+0.16}_{-0.26}$ & $-0.83^{+0.18}_{-0.37}$ \\
   \hline
   $8.0$ & $-0.65^{+0.18}_{-0.29}$ & $-1.12^{+0.22}_{-0.40}$ \\
   \hline
   $8.5$ & $-1.01^{+0.21}_{-0.32}$ & $-1.69^{+0.31}_{-0.45}$ \\
   \hline
   $9.0$ & $-2.02^{+0.34}_{-0.38}$ & $-3.29^{+0.57}_{-0.56}$ \\
   \hline
 \end{tabular}
 \end{center}
 \end{table}

The uncertainties shown in Fig.~\ref{fig:timl} are estimated
by varying each input parameter involved in the
calculation of $\Delta C_9^{BK}(q^2<0)$ and
then recalculating   $\Delta C_9^{BK}(q^2>0)$  from the dispersion relation
with the  modified input.  The resulting individual deviations
of     $\Delta C_9^{BK}(q^2>0)$ are then added in quadrature.
Furthermore, we repeat the fit of the dispersion relations
for the three alternative parametrizations of the dispersion integrals  (I)-(III)
specified at the end of the previous section.
We interpret
the difference between $\Delta C_9^{BK}(q^2>0) $ obtained
with the default parametrization and the one in  the model III
(which only contains $J/\psi$ and $\psi(2S)$ resonances)  as the ``systematic'' uncertainty of our approach and include it in the uncertainty budget.
As an example, we present the correction to $C_9$
calculated at one particular value of dilepton mass:
\ba
{\rm Re}[\Delta C_9^{BK}(q^2=4 \,{\rm GeV}^2)]&=&-0.21
\,{}^{+0.03}_{-0.04}
\bigg|_{\lambda_B}
\,{}^{+0.00}_{-0.10} \bigg|_{a_1^K}
\,{}^{+0.08}_{-0.00} \bigg|_{\mu}
\,{}^{+0.02}_{-0.01} \bigg|_{f_{BK}^{+}(0)}
\,{}^{+0.02}_{-0.06} \bigg|_{syst.} \,,
\nonumber   \\
\\
{\rm Im}[\Delta C_9^{BK}(q^2=4 \,{\rm GeV}^2)]&=&-0.34
\,{}^{+0.05}_{-0.07}
\bigg|_{\lambda_B}
\,{}^{+0.01}_{-0.12} \bigg|_{a_1^K}
\,{}^{+0.08}_{-0.07} \bigg|_{a_2^K}
\,{}^{+0.06}_{-0.03} \bigg|_{\mu}
\,{}^{+0.05}_{-0.02} \bigg|_{f_{BK}^{+}(0)}
\,{}^{+0.02}_{-0.04} \bigg|_{syst.}, \nonumber
\label{eq:deltaCuncert}
\ea
where all significant ($ >\pm 5\%$) uncertainties  are shown separately.
As expected, the  dependence on the $B\to K$ form factor is inessential.
Here,  we assume a fixed uncertainty for
$B\to K$ form factors  stemming from LCSR and
neglect possible correlations when varying  the rest of the input.
For convenience, we also present  the numerical intervals for $\Delta C_9^{BK}(q^2)$
in the region $1.5 \, {\rm GeV^2} < q^2 < 9.0 \, {\rm GeV^2}$  in Table~\ref{tab:res}.
They have to be compared with $C_9\simeq 4.0-4.5$
(see App.~A).  The magnitude of $\Delta C_9^{BK}(q^2)$
remains in the ballpark of $\sim 10 \%$  of $C_9$ at small and intermediate $q^2$
and grows approaching the charmonium region.
We refrain from quoting $\Delta C_9^{BK}(q^2)$ in the region between
$J/\psi$ and $\psi(2S)$ where the ``systematic'' uncertainty caused by the
parameterization of dispersion relation is rather large.

\section{ Observables in $B\to K \ell^+\ell^-$ }

Substituting  the numerical results for  ${\cal H}^{(BK)}(q^2)$ to
the amplitude  (\ref{eq:ampl}), we calculate
 the differential width of $\bar{B}^0\to \bar{K}^0 \mu^+\mu^-$
in the region $4m_\mu^2<q^2<4m_D^2$  (see Fig.~\ref{fig:BR}).
Integrating over the typical intervals of dilepton-mass squared
(bins) selected in the experiments,  we present the resulting
partial widths  in Table~\ref{tab:BRbins}, in comparison
with the available measurements. In order to illustrate the
origin of the quoted uncertainties, we present them separately for one
of the bins:
\ba
\int\limits_{1.0 ~{\rm GeV}^2}^{6.0 ~{\rm GeV}^2} dq^2
\frac{dBR(\bar{B}^0\to \bar{K}^0\mu^+\mu^-)}{dq^2}
\nonumber\\
=\Big(
1.76 \,\,{}^{+0.58}_{-0.21} \bigg|_{f_{BK}^{+}(0)}
\,\,{}^{+0.16}_{-0.09} \bigg|_{slope}
\,\,{}^{+0.01}_{-0.01} \bigg|_{\mu}
\,\,{}^{+0.01}_{-0.02} \bigg|_{syst.}
\Big)\times 10^{-7}\,,
\label{eq:deltB}
\ea
where the remaining
individual uncertainties smaller than 1\% are not shown.
As opposed to $\Delta C_9^{(BK)}$,  the theoretical uncertainty of the width
mainly originates from the normalization of the form factor  $f_{BK}^+$.
Our predictions for the partial widths of
$\bar{B}^0\to \bar{K}^0\ell^+\ell^-$ in Table~\ref{tab:BRbins}
are somewhat larger than the  results of the most recent LHCb measurement
\cite{LHCbnew}  of  $B^-\to K^- \ell^+\ell^-$.
\begin{figure}[tb]
\begin{center}
\hspace{-1 cm}
\includegraphics[scale=0.9]{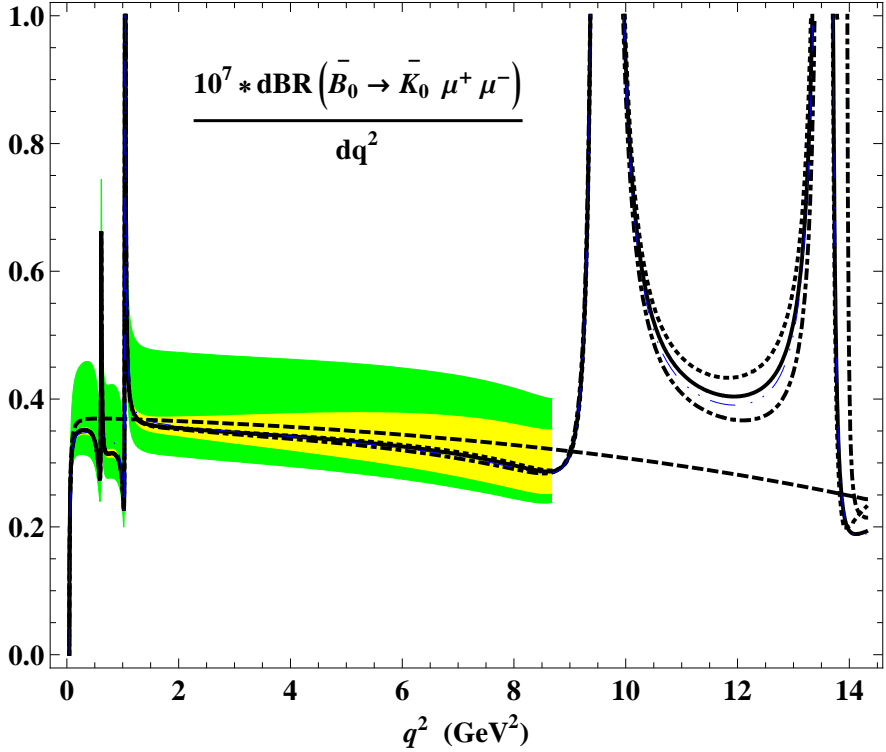}
\caption{ \it Differential branching fraction of  $\bar{B}^0\to \bar{K}^0 \mu^+\mu^-$.
The solid line corresponds to the central input and to the default parametrization
for the dispersion integrals. The darker (green) and brighter (yellow) shaded area
indicates the uncertainties  including (excluding) the one from the form
factor normalization.
The alternative parametrizations I, II and III of the dispersion integrals
yield dotted, dash-dotted and thin dash-double-dotted (blue) lines,
respectively. The long-dashed line corresponds
to the width calculated without nonlocal hadronic effects. }
\label{fig:BR}
\end{center}
\end{figure}
\begin{table}[h]
\caption{ $d{\rm BR}(B\to K \mu^+ \mu^-)/dq^2$
integrated over $[q_{min}^2,q_{max}^2]$ in units of  $10^{-7}$. }
\begin{center}
{\small
\begin{tabular}{|c|c|c|c|c|c|c|c|}
\hline
 &&&&& \\
  $ [q_{min}^2,q_{max}^2] $    &  Belle \cite{Wei:2009zv}   &
CDF  \cite{Aaltonen:2011qs}  &  LHCb  \cite{LHCbBKisosp} & LHCb  \cite{LHCbnew} &  this
work          \\
 ${\rm  GeV^{2}}$  &&&&& \\ \hline
   &&&&& \\
$[0.05, 2.0]$   &  $0.81^{+0.18}_{-0.16} \pm
0.05$ & $ 0.33 \pm 0.10 \pm 0.02$   & $0.21^{+0.27}_{-0.23}$ & $0.56 \pm 0.05 \pm 0.03$ &
$0.71^{+0.22}_{-0.08}$         \\
&&&&&  \\ \hline
&&&&&  \\
$[2.0, 4.3]$      &  $0.46^{+0.14}_{-0.12} \pm
0.03$ & $0.77 \pm 0.14 \pm 0.05$   & $0.07^{+0.25}_{-0.21}$ & $0.57 \pm 0.05 \pm 0.02$ & $
0.80^{+0.27}_{-0.11}$       \\
&&&&&  \\ \hline
&&&&& \\
$[4.3, 8.68]$  &  $1.00^{+0.19}_{-0.08}\pm 0.06$       &
$1.05\pm0.17\pm 0.07$   & $1.2 \pm 0.3$ & $1.00 \pm 0.07 \pm 0.04$ & $1.39^{+0.53}_{-0.22}$ \\
&&&&& \\ \hline
&&&&& \\
$[1.0, 6.0] $     &  $1.36^{+0.23}_{-0.21} \pm
0.08$ & $1.29 \pm 0.18 \pm 0.08$  & $0.65^{+0.45}_{-0.35}$ & $1.21 \pm 0.09 \pm 0.07$  &
$1.76^{+0.60}_{-0.23}$\\
&&&&& \\  \hline
\end{tabular}
}
\label{tab:BRbins}
\end{center}
\end{table}

\begin{table}[h]
\caption{ Isospin asymmetry
calculated from the partial widths of
$\bar{B}^0\to \bar{K}^0\ell^+\ell^-$ and  $B^-\to K^-\ell^+\ell^-$
integrated over $1.0<q^2<6.0$ GeV$^2$.}
\begin{center}
  \begin{tabular}{|c|c|c|c|c|}
\hline
 &&& \\
 Belle \cite{Wei:2009zv}   &  BaBar \cite{Lees:2012vw}  &  LHCb
\cite{LHCbBKisosp}  &  this work  \\[1mm]
\hline
&&& \\[-1mm]
$-0.41^{+0.25}_{-0.20} \pm
0.07$ & $ -0.41 \pm 0.25 \pm 0.01$  & $-0.35^{+0.23}_{-0.27}$    &
$(-0.4)\% \div (-0.3)\% $        \\
&&& \\  \hline
\end{tabular}
\label{tab:bins of isospin}
\end{center}
\end{table}

In addition, we also calculated the  (CP averaged) isospin asymmetry  in
$B\to K \ell^+\ell^-$   defined as
\begin{eqnarray}
a_{I}^{(0-)}(q^2)  ={ d \Gamma(\bar{B}_0 \to
\bar{K}_0 \ell^+\ell^-)  / d q^2 - d \Gamma( B^{-} \to K^{-}
\ell^+\ell^-) / d q^2  \over  d \Gamma(\bar{B}_0 \to \bar{K}_0
\ell^+\ell^-)  / d q^2 + d \Gamma( B^{-} \to K^{-} \ell^+\ell^-) /
d q^2  } \,,
\label{eq:isoas}
\end{eqnarray}
In our approach this effect originates from the small  differences
between the amplitudes of photon emission from the $d$ and $u$ spectator quarks, in the
nonfactorizable NLO contributions and in the weak annihilation.
The result for the differential asymmetry  defined above is
shown in Fig. \ref{fig:isospin} and does not  exceed a $\pm 4 \%$ level within estimated errors.
At $q^2> 2 $ GeV$^2$ our expectation for the
isospin asymmetry  in $B \to K \ell^+\ell^-$  is in the ballpark of the prediction
obtained in \cite{Feldmann:2002iw} for the isospin asymmetry in $B\to K^*\ell^+\ell^-$.
In  Table \ref{tab:bins of isospin}  we present the integrated
isospin asymmetry defined similarly to (\ref{eq:isoas}) but with the partially integrated
widths instead of differential widths. This integrated
characteristics is expected to be very small,
revealing an intriguing tension  with the available experimental measurements
presented in the same table. It is important to further improve the
accuracy of our prediction. E.g., the calculation presented here
assumes isospin symmetry for the $B\to K$ form factors. This issue
 deserves a separate study but  -- having in mind
the usual magnitude of  isospin violation -- it is hard to expect that the deviation
of $f^+_{B^0K^0}$ from  $f^+_{B^-K^-}$ brings
substantial changes in the isospin asymmetry  of $B \to K \ell^+\ell^-$.
\begin{figure}[t]
\begin{center}
\hspace{-1 cm}
\includegraphics[scale=0.6]{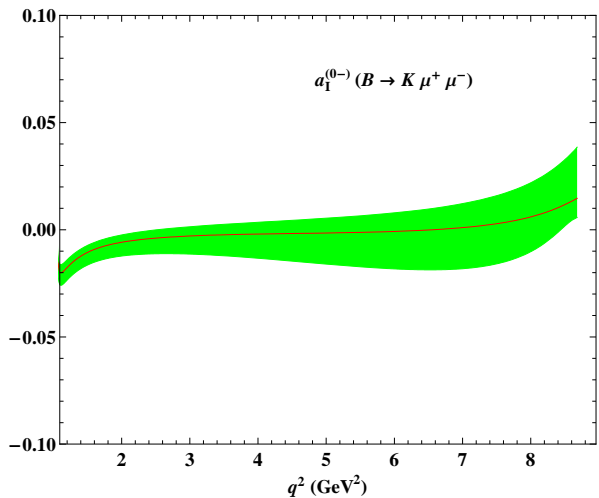}
\caption{Isospin asymmetry $a_{I}^{(0-)}(q^2)$ in
$B\to K \ell^+\ell^-$ (the solid curve), the shaded (green) band indicates
the uncertainty of
our prediction.}
\label{fig:isospin}
\end{center}
\end{figure}

Having at hand the SM prediction for the amplitude of $B\to K \ell^+\ell^-$
it is interesting to investigate the potential influence of
new physics in $b\to s \ell^+\ell^-$  on the observables in this decay.
Recent improvement \cite{LHCbBmumu}  of the upper bound for
$BR(B_{s} \to \mu^{+} \mu^{-})$ has already put substantial constraints on
the new FCNC operators involving pseudoscalar and
scalar couplings of the lepton pair.  Hence
we only consider
a generic new physics  induced by the two tensor operators:
\begin{eqnarray}
O_T= \frac{\alpha_{em}}{2\pi}\left(\bar{s}   \sigma_{\mu \nu} b
\right) \left( \bar{l}  \sigma^{\mu \nu} l \right)\,, \qquad
O_{T5}= \frac{\alpha_{em}}{2\pi}\left(\bar{s}   \sigma_{\mu \nu} b
\right) \left( \bar{l}  \sigma^{\mu \nu} \gamma_5 l \right) \,,
\label{eq:OT}
\end{eqnarray}
with Wilson coefficients $C_T$ and $C_{T5}$, respectively.
\begin{figure}[t]
\begin{center}
\hspace{-1 cm}
\includegraphics[scale=0.6]{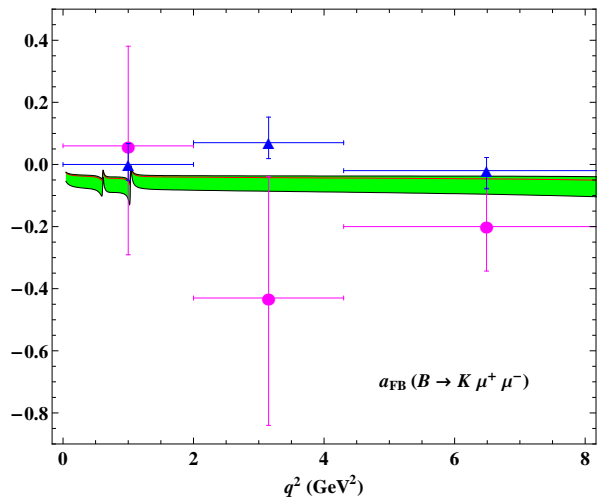} \hspace{0.2  cm}
\caption{Forward-backward asymmetry in $B \to K  \ell^+\ell^-$
emerging due to new physics contributions of tensor
operators with  the Wilson coefficients
$C_T= C_{T5}=1.2$. The shaded (green) band indicates the uncertainty
of our prediction due to variation of SM parameters. The data points are from the
measurement by LHCb \cite{LHCbnew} (triangles)  and Belle \cite{Wei:2009zv} (circles). }
\label{fig:AFB}
\end{center}
\end{figure}

The $B \to K  \ell^+\ell^-$ width
is   calculated adding the new operator contributions to the decay amplitude and
assuming $C_T=\pm C_{T5}$.
As shown in \cite{Bobeth:2007dw}, these two parameters are effectively constrained by
comparing the measured upper bounds  on the inclusive branching fraction
$\bar{B} \to X_s \ell^+\ell^-$  with the SM prediction.
Typically one obtains $|C_T|\sim  |C_{T5}|<1.2$.
With this constraint, our result for  the decay rate of
$B\to K \ell^+\ell^-$, including the new physics contribution
and integrated over the interval $1.0<q^2<6.0$ GeV$^2$,
reveals a rather small deviation from the SM  prediction, at the level of $< 5\%$.
The  predicted forward-backward asymmetry in $B\to K \ell^+\ell^-$,
emerging due to the  new operators (\ref{eq:OT}),
is plotted in  Fig. \ref{fig:AFB} at  $C_T=C_{T5}=1.2$.
This observable can reach at most $- (5-10)\%$, being  almost independent of $q^2$.
There are measurements of the forward-backward asymmetry in $B\to K \ell^+\ell^-$
by Belle Collaboration \cite{Wei:2009zv} and more recently by LHCb Collaboration \cite{LHCbnew},
see Fig. \ref{fig:AFB}. However, these measurements are not yet sensitive to such small effects.

\section{Discussion}

In this paper, we calculated the  nonlocal hadronic contributions
to $B\to K  \ell^+\ell^-$ generated by the four-quark and gluon-penguin operators
of the effective Hamiltonian, combined with the e.m. emission
of the lepton pair.
In this calculation the LCSR results
for the form factors and for the soft-gluon nonfactorizable effects
are used, whereas the hard-gluon effects are approximated
employing QCD factorization.
We followed the method suggested in \cite{KMPW}, combining the
QCD results for the nonlocal hadronic matrix elements
valid in the region of negative dilepton-mass squared, with the
dispersion relations  in the physical region.
In these relations the residues of the lowest vector mesons
are fixed from the data, whereas a nontrivial destructive interference
between the dominant $J/\psi$ and $\psi(2S)$ contributions
plays an important role. A model ansatz is adopted for the higher state
contributions to the dispersion relations,
with the parameters fitted to the nonlocal amplitude calculated in QCD.

Our main result is the numerical prediction for the nonlocal hadronic
amplitude ${\cal H}^{(BK)}(q^2)$ cast in the form of the correction
to the Wilson coefficient $C_9$. This correction displayed
in Table~\ref{tab:res} and in Fig.~\ref{fig:timl} can be used in future
phenomenological analysis  of the $B \to K  \ell^+\ell^-$ decay.
Employing dispersion relation, we avoid
unphysical ``kinks'' in the observables generated by the quark-antiquark
production thresholds which appear if one directly uses QCD diagrams in the
physical region (see e.g., \cite{Ali:1999mm}). Probing alternative
parametrizations  of the dispersion integrals we argue that
the ``systematic'' uncertainty introduced by this procedure
is inessential in the large hadronic recoil region,
practically up to $q^2=m^2_{J/\psi}$. The approach used here
is not applicable at larger values of
dilepton mass, especially at $q^2>4m_D^2$, where a complicated
interference of charmonium resonances and the proximity of the singularities
of the $B\to K$ form factors brings too many hadronic degrees of freedom into play
\footnote{ Let us mention that in the low hadronic-recoil region a
different method is used \cite{lowrecoil}, based on the local OPE of the nonlocal
hadronic amplitudes valid at $q^2\sim m_b^2\to \infty$. Applying this method
to $B \to K^{(*)}  \ell^+\ell^-$
decay, one relies on the (quasi)local quark-hadron duality and
smallness of power corrections. The connection of the local OPE
to the approach  used in this paper is an
interesting open problem that deserves a dedicated study.}.

In the large hadronic recoil region, we evaluated the
partial width and isospin asymmetry of $B \to K  \ell^+\ell^-$.
The impact of nonlocal hadronic contributions on the width
is moderate below the charmonium region.
The main uncertainty of our prediction in this region stems
from  the normalization of the $B\to K$ form factors,
leaving a room  for improving the accuracy of the SM  prediction for $B \to K  \ell^+\ell^-$. In particular, new precise lattice
calculations of these form factors combined with
updated LCSR results at small $q^2$ are desirable.
In addition, we predict a very small  isospin asymmetry in $B \to K  \ell^+\ell^-$,
hence, more precise measurements of this observable are needed.

A further improvement of our prediction for the effective correction $\Delta C_9^{(BK)}(q^2)$
is possible along two lines: firstly, achieving more accuracy in QCD calculations
at negative $q^2$ and secondly, with more detailed parametrizations of the dispersion
integrals adding the excited vector meson terms. In this respect, dedicated measurements
of nonleptonic decays of the type $B\to V' K$ where $V'$ are radial excitations
of $\rho,\omega, \phi$ and the charmonium levels starting from $\psi(3772)$
will be helpful.

The study presented in this paper
could be extended to $B\to K^*\ell^+
\ell^-$, for which the charm-loop effect has already been analyzed
in \cite{KMPW}. However, calculating the other effects
in this process  will be more demanding.  E.g., the nonlocal
hadronic contributions proportional to $Q_s$
such as the $\phi$-meson pole will be
enhanced at small $q^2$, being multiplied by
the factor $1/q^2$ from the virtual photon propagator in the
amplitude  for the transversely polarized $K^{\ast}$ meson.
To accurately analyze this particular contribution,
a separation of $Q_s-Q_b$ component from the $Q_c$-one
is desirable in all three nonlocal
hadronic amplitudes ${\cal H}^{(BK^*)}_{1,2,3}$
corresponding to the three polarization states of the $K^*$ meson.
This demands a dedicated ``flavour splitting''  of the NLO diagrams
calculated in \cite{Asatrian:2001de, Asatryan:2001zw}
on one hand and a separate dispersion
relation with $\bar{s}s$ vector-meson states on the other hand.
Also the NLO two-loop diagrams induced
by  the penguin operators should be included
to achieve an adequate accuracy.
Altogether, the QCD calculation at negative $q^2$ and
a complete analysis of hadronic dispersion relations
for  all three invariant amplitudes of $B\to K^*  \ell^+\ell^-$
represents a significantly more challenging task. The
nonlocal corrections to $C_9$
are expected to be numerically larger than for the kaon mode,
as already seen from the comparison of charm loop effects
for both processes in \cite{KMPW}.
Hence, a substantial impact of the nonlocal hadronic effects
on  $B\to K^*  \ell^+\ell^-$ is anticipated, even in the large hadronic recoil
region.
Last but not least, as already discussed in the introduction,
the fact that $K^*$ has a finite width decaying to  $K \pi$,
hinders one from the calculation of ``pure'' $B\to K^*$ form factors, with the
accuracy achieved for the $B\to K$ form factors,
either on the lattice or with QCD sum rules.
New approaches are desirable to calculate the  form factors
of $B$-meson transitions to the two-meson system including resonances.
Only in this case the theory can meet the challenge of continuously
improving experimental measurements  of exclusive FCNC $B$ decays.

\section*{Acknowledgements}

We are thankful to H. Asatrian and C. Greub for useful discussions
on their NLO diagram calculation. This work is supported by the German
Ministry of Research (BMBF), Contract No. 05H09PSF.

\section*{Appendix A: Effective Hamiltonian}

Here we list  the operators entering the effective Hamiltonian (\ref{eq:Heff}):
\begin{eqnarray}
 O_1=
\left(\bar{s}_L\gamma_\rho c_L\right) \left(\bar{c}_L\gamma^\rho
b_L\right)\,, & \qquad &
~~~ O_2 =
\left(\bar{s}^j_{L}\gamma_\rho c^i_{L}\right) \left(\bar{c}^i_{L}\gamma^\rho
b^j_{L}\right)\,, \nonumber
\\
O_3 = \left(\bar{s}_L\gamma_\rho b_L\right) \sum_q
\left(\bar{q}_L\gamma^\rho q_L\right)\,, & \qquad &   ~~~
O_4 = \left(\bar{s}^{i}_L \gamma_\rho b^{j}_L\right) \sum_q
\left(\bar{q}^{j}_L\gamma^\rho q^{i}_L\right) \,, \nonumber
\\
O_5 = \left(\bar{s}_L\gamma_\rho b_L\right) \sum_q
\left(\bar{q}_R\gamma^\rho q_R\right)\,, & \qquad &  
O_6 = \left(\bar{s}^{i}_L \gamma_\rho b^{j}_L\right) \sum_q
\left(\bar{q}^{j}_R\gamma^\rho q^{i}_R\right) \,,  \nonumber\\
O_{7\gamma}=-{e \over 16 \pi^2} \bar{s} \sigma_{\mu \nu} (m_s L+
m_b R)b F^{\mu \nu },
& \qquad &
O_{8 g}=-{g_s \over 16
\pi^2} \bar{s}_{i} \sigma_{\mu \nu} (m_s L+ m_b R) T^a_{ij}
b_{j} G^{a \mu \nu }\, , \nonumber
\\
~~~
O_9= \frac{\alpha_{em}}{4\pi}\left(\bar{s}_L\gamma_\rho b_L\right) \left(
\bar{l}\gamma^{\rho}l \right)\,,
& \qquad &
O_{10}=\frac{\alpha_{em}}{4\pi}\left(\bar{s}_L\gamma_\rho b_L\right)
\left(\bar{l}\gamma^{\rho}\gamma_5 l \right)\, , \nonumber
\label{eq:listoper}
\end{eqnarray}
where the notations are $q_{L(R)}=\frac{1-(+)\gamma_5}{2}q$ and
$L(R)=\frac{1-(+)\gamma_5}{2}$ .
 We use the standard conventions for the operators
$O_i$, except the labeling of $O_1$ and $O_2$ is interchanged.
The sign convention for $O_{7 \gamma}$ and $O_{8 g}$ corresponds
to the covariant derivative $iD_\mu= i\partial_\mu+eQ_f A_{\mu} + g T^a
A^a_{\mu}$, where $Q_f$ is the fermion charge. In addition, the convention for
the Levi-Civita tensor adopted in this work is
${\rm Tr}\{\gamma^\mu\gamma^\nu\gamma^\rho\gamma^\lambda\gamma^5\}=4i
\epsilon^{\mu\nu\rho\lambda} \,,~ \epsilon^{0123}=-1.$
The numerical values of the  Wilson coefficients
at three different values of the scale $\mu$ are
listed in the table below \cite{BBL}:

\begin{center}
 \begin{tabular}{|c|c|c|c|c|}
  \hline
  $\mu$ (GeV)                 &  $0.5 m_b$    &  $ m_b$       & $1.5 m_b$          \\  \hline
   $C_1$                      &  $1.193$       &  $ 1.117$      & $ 1.090$         \\  \hline
   $C_2$                      &  $-0.401$       &  $-0.267$      & $ -0.214$       \\  \hline
   $C_3$($\times 10^{-2}$)    &  $1.919$       &  $1.206$      & $0.931$         \\  \hline
   $C_4$($\times 10^{-2}$)    &  $-3.964$      &  $-2.750$     & $-2.225$      \\  \hline
   $C_5$($\times 10^{-2}$)    &  $1.041$      &  $0.770$     & $0.639$     \\  \hline
   $C_6$($\times 10^{-2}$)    &  $-5.479$      &  $-3.411$     & $-2.637$    \\  \hline
   $C_{7}$                    &  $-0.370$       &  $-0.320$      & $-0.297$ \\
\hline
     $C_{8g}$                    &  $-0.184$       &  $-0.166$      & $-0.157$ \\
\hline
$C_9$                      &  $4.450$      &  $4.232$     & $4.029$     \\  \hline
   $C_{10}$                   &  $-4.410$      &  $-4.410$      & $-4.410$        \\  \hline
    \end{tabular}
\end{center}
We also use
$ |V_{ts}|=0.0405^{+0.0006}_ {-0.0010} $ ,  $|V_{tb}|=0.999^{+0.000043}_{ -0.000025}$
\cite{CKMfit}  and  $\alpha_{em}=1/129$.

\section*{Appendix B: $B$ -Meson Light-cone DA's}

In $x^2=0$ limit we adopt the following
decomposition of  the $B$-to-vacuum matrix element into
four independent  three-particle DA's (see e.g., \cite{KKQT}):
\begin{eqnarray}
&&\langle 0|\bar{q}_\alpha(x) G_{\lambda\rho}(ux)
h_{v\beta}(0)|\bar{B}^0(v)\rangle=
\frac{f_Bm_B}{4}\int\limits_0^\infty d\omega
\int\limits_0^\infty d\xi\,  e^{-i(\omega+u\xi) v\cdot x}
\nonumber \\
&&\times \Bigg [(1 +\DS v) \Bigg \{ (v_\lambda\gamma_\rho-v_\rho\gamma_\lambda)
\Big(\Psi_A(\omega,\xi)-\Psi_V(\omega,\xi)\Big)
-i\sigma_{\lambda\rho}\Psi_V(\omega,\xi)
\nonumber\\
&&-\left(\frac{x_\lambda v_\rho-x_\rho v_\lambda}{v\cdot x}\right)X_A(\omega,\xi)
+\left(\frac{x_\lambda \gamma_\rho-x_\rho \gamma_\lambda}{v\cdot x}\right)Y_A(\omega,\xi)\Bigg\}\gamma_5\Bigg]_{\beta\alpha}\,,
\label{eq-B3DAdef}
\end{eqnarray}
where the path-ordered gauge factors on l.h.s. are omitted for
brevity.  For the three-particle $B$-meson
DA's the model suggested in \cite{KMO2} is used:
\ba
 \Psi_A(\omega,\xi)&=&\Psi_V(\omega,\xi)= { \lambda_E^2 \over 6
 \omega_0^4} \xi^2 e^{-(\omega+\xi)/ \omega_0} \, , ~~
 X_A(\omega,\xi) = { \lambda_E^2 \over 6 \omega_0^4} \xi (2
 \omega -\xi) e^{-(\omega+\xi) /\omega_0} \, , \nonumber \\
 Y_A(\omega,\xi) &=& - { \lambda_E^2 \over 24 \omega_0^4} \xi (7
 \omega_0 -13 \omega + 3\xi) e^{-(\omega+\xi) /\omega_0} \,.
 \label{eq:3partDA}
 \ea
In this model the parameter $\omega_0$ is equal to the
inverse moment $\lambda_B$ of the $B$ meson two-particle DA
$\phi_B^{+}$. For the latter we use  the  ansatz
suggested in \cite{GN}:
\begin{eqnarray}
\phi^+_B(\omega)  =  \dfrac{\omega}{\omega_0^2}\,e^{-\frac{\omega}{\omega_0}}\,,
~~~
\phi^-_B(\omega) =  \dfrac{1}{\omega_0}\,e^{-\frac{\omega}{\omega_0}}\,.
\label{eq-GN}
\end{eqnarray}

\section*{Appendix C: Functions entering LCSR (\ref{eq:QCDcorr}) }

Here we collect the expressions for  the coefficient
functions $F_{n}^{(DA)}(q^2,\omega)$
entering the answer for the correlation function in (\ref{eq:QCDcorr})
and used to derive the LCSR  (\ref{eq:SRO8}).
 Below  the short-hand notation
$\sigma=\omega/m_B, \bar{\sigma}=1-\sigma$ is introduced:

\begin{eqnarray}
 F_{1}^{(\Psi A)}(q^2,\omega)
=-{3 Q_b \over m_b^2}
  (m_s-m_B \bar{\sigma}) \left( {2 m_b \over m_B}- 1 \right)
-{3 Q_s \over  \bar{\sigma}^2 m_B^2-m_s^2} (2 \sigma-1)
 (m_s-m_B \bar{\sigma}) \,, \nonumber
\end{eqnarray}
\begin{eqnarray}
F_{1}^{(\Psi V)}(q^2,\omega)
={3 Q_b \over m_b^2}
  (m_s+m_B \bar{\sigma}) \left( {2 m_b \over m_B}-1 \right)
+ {3 Q_s \over  \bar{\sigma}^2 m_B^2-m_s^2} (2 \sigma-1)
 (m_s+m_B \bar{\sigma}) \,, \nonumber
\end{eqnarray}
\begin{eqnarray}
 F_{1}^{(XA)}(q^2,\omega)
=-{Q_b \over m_b^2}
 \bigg[ \left(2+{m_b \over m_B} \right)  - {m_s \over \bar{\sigma} m_B} \left(1+ {2m_b \over
 m_B} \right)  \bigg ]
\nonumber \\
- { Q_s \over  \bar{\sigma}^2 m_B^2-m_s^2} \left( 1- {m_s \over \bar{\sigma} m_B} \right)  
- {3 m_B^2 Q_s \over  ( \bar{\sigma}^2 m_B^2-m_s^2 )^2} (2 \sigma-1)
  \left({ m_s^2 \over m_B^2} - \bar{\sigma}^2 \right)\,, \nonumber
\end{eqnarray}
\begin{eqnarray}
\nonumber \\
 F_{1}^{(YA)}(q^2,\omega)
=-{4 Q_b \over m_b^2}
 \left(1-{m_b \over m_B} \right)  - { 4 Q_s \over  \bar{\sigma}^2 m_B^2-m_s^2}
- {6 m_s m_B Q_s \over  ( \bar{\sigma}^2 m_B^2-m_s^2 )^2} (2 \sigma-1)
 \left( {m_s \over m_B} - \bar{\sigma}  \right) \,,
\nonumber
 \end{eqnarray}

\begin{eqnarray}
 F_{2}^{(\Psi A)}(q^2,\omega)
={F}_{2}^{(\Psi V)}(q^2,\omega)=0 \,,
\nonumber
\end{eqnarray}
\begin{eqnarray}
F_{2}^{(X A)}(q^2,\omega)
=-{Q_b \over m_b^2}
 \bigg \{\left( \bar{\sigma} m_B^2 -{m_s^2 \over \bar{\sigma} }
 \right)  \bigg [ \left ( {m_s \over m_B}+\bar{\sigma} \right )
 +{2 m_b \over m_B} \left({m_s \over m_B} - \bar{\sigma} \right)  \bigg ] \nonumber \\
-  q^2  \bigg[ 1 - {2 m_b \over \bar{\sigma} m_B}  \left
(\bar{\sigma} - {m_s \over m_B} \right )
 +  {m_s \over m_B}  { 4 \sigma-3 \over \bar{\sigma}}  \bigg]  \bigg \}
\nonumber \\
+ { Q_s \over  \bar{\sigma}^2 m_B^2-m_s^2} \left( 1- {m_s \over \bar{\sigma} m_B} \right)
\bigg \{ q^2(2\sigma-1)
+m_B^2\bigg [ \bar{\sigma}^2 (2 \sigma+1) +  {m_s \over m_B} \left
(2 \bar{\sigma} +(1-2 \sigma) {m_s \over m_B} \right) \bigg ]
\bigg \}\,, \nonumber
\end{eqnarray}

\begin{eqnarray}
F_{2}^{(Y A)}(q^2,\omega)
=-{2 Q_b \over m_b^2}
 \bigg \{  2 q^2  \left( \bar{\sigma} - { m_b \over  m_B} \right)
\nonumber \\
+   m_B \left(  m_s- \bar{\sigma} m_B  \right)  \bigg [ 2
\bar{\sigma} (1- {m_b \over m_B})
-{m_s \over m_B}(1-4{m_b \over
m_B})  \bigg ] \bigg \}  \nonumber \\
+ { 2 Q_s \over  \bar{\sigma}^2 m_B^2-m_s^2}
\bigg \{ m_B^2 \left( \bar{\sigma} - { m_s \over  m_B} \right)
\bigg [2 \bar{\sigma}^2 +  { m_s \over  m_B}(4 \sigma-1) \bigg ]
- 2q^2 (1-2 \sigma) \bigg \}\,.
\end{eqnarray}

Finally, we present the substitution relations for the integrals in
(\ref{eq:QCDcorr}):
\begin{eqnarray}
\int_0^{\infty} d \omega
\frac{f_{1}(q^2,\omega)}{ (p-\omega v)^2-m_s^2}  \to
- \int_0^{\omega_0} \frac{ d \omega}{  1-\omega /m_B}
f _1(q^2,\omega)\, e^{-s/M^2}  \,,
\nonumber \\[4mm]
\int_0^{\infty} d \omega
\frac{f_{2}(q^2,\omega)}{ [(p-\omega v)^2-m_s^2]^2 }
\to
\int_0^{\omega_0} { d \omega \over (1-\omega /m_B)^2}
\frac{f_{2}(q^2,\omega)}{M^2}e^{-s/M^2}
\nonumber \\
+ {\eta(\omega_0) \over m_B} {e^{-s_0/M^2} \over (1-\omega_0 /m_B)^2} f_ {2}
(q^2,\omega_0) \,,
\end{eqnarray}
where $f_{1,2}$  is the product of the function $F_{1,2}^{(DA)}$
and the corresponding DA  (the latter integrated over $\xi$), and
\begin{eqnarray}
s &=& \omega m_B +{ m_s ^2 -  q^2 \omega /m_B \over
1-\omega/m_B}\,,
 \qquad  \eta(\omega_0) = \bigg[ 1+ {m_s^2 -q^2 \over
(m_B-\omega_0)^2 } \bigg]^{-1} \,,
\nonumber \\
\omega_0 &=& { (s_0 + m_B^2-q^2) - \sqrt{(s_0 + m_B^2-q^2)^2 + 4 m_B^2
(m_s^2-s_0)} \over 2 m_B} \,. \nonumber
\end{eqnarray}


\end{document}